\documentclass{aa}

\newcommand\gcr[1]{{#1}}

\usepackage{amsmath}
\usepackage{flexisym}
\usepackage{graphicx}

\usepackage{booktabs}
\usepackage{tabularx}
\usepackage[breaklinks]{hyperref}
\usepackage{enumitem}

\def \micron {{$\mu \rm m$}}

\def \Lsixum {{{\rm L}_{6\mu \rm  m}}}

\def \Msun {{$\rm M_{\odot}$}}
\def \ergs {{$\rm ergs \, s^{-1}$}}

\usepackage{graphicx}
\usepackage{txfonts}
\begin{document}

\title{The multiwavelength properties of red QSOs - Evidence for dusty winds as the origin of QSO reddening}
\titlerunning{ Evidence for dusty winds as the origin of QSO reddening}


\author{G. Calistro Rivera \inst{1}\thanks{ESO Fellow} \and D. M. Alexander\inst{2} \and  D. J. Rosario \inst{2} \and  C. M. Harrison \inst{3} \and M. Stalevski \inst{4,5} \and S. Rakshit \inst{6,7} \and V. A. Fawcett  \inst{2}  \and L.~K. Morabito  \inst{2} \and L. Klindt  \inst{2}  \and P. N. Best \inst{8} \and M. Bonato \inst{9,10,11} \and R. A. A. Bowler \inst{12} \and T. Costa \inst{13} \and R.  Kondapally  \inst{8}   
}%

\institute{European Southern Observatory (ESO), Karl-Schwarzschild-Stra\ss e 2, 85748 Garching bei M\"unchen, Germany \\\email{gabriela.calistrorivera@eso.org} 
\and
Centre for Extragalactic Astronomy, Department of Physics, Durham University, Durham DH1 3LE, UK
\and
School of Mathematics, Statistics and Physics, Newcastle University, Newcastle upon Tyne, NE1 7RU, UK
\and 
Astronomical Observatory, Volgina 7, Belgrade 11060, Serbia
\and
Sterrenkundig Observatorium, Universiteit Gent, Krijgslaan 281, Gent B-9000, Belgium
\and
Finnish Centre for Astronomy with ESO (FINCA), University of Turku, Quantum, Vesilinnantie 5, 20014, Finland
\and
Aryabhatta Research Institute of Observational Sciences, Manora Peak, Nainital 263002, India
\and
Institute for Astronomy, University of Edinburgh, Royal Observatory, Blackford Hill, Edinburgh EH9 3HJ, UK
\and
INAF-Istituto di Radioastronomia, Via Gobetti 101, I-40129, Bologna, Italy
\and
Italian ALMA Regional Centre, Via Gobetti 101, I-40129, Bologna, Italy
\and
INAF-Osservatorio Astronomico di Padova, Vicolo dell'Osservatorio 5, I-35122, Padova, Italy.
\and 
Department of Astrophysics, University of Oxford, The Denys Wilkinson Building, Keble Road, Oxford OX1 3RH, UK
\and 
Max-Planck-Institut f\"ur Astrophysik, Karl-Schwarzschild-Str\ss e 1, D-85748 Garching bei M\"unchen, Germany
}

  \abstract
  {
Fundamental differences in the radio properties of red quasars (QSOs), as compared to blue QSOs, have been recently discovered, positioning them as a potential key population in the  evolution of galaxies and black holes across cosmic time. 
To elucidate the nature of these objects, we exploited a rich compilation of broad-band photometry and spectroscopic data to model their spectral energy distributions (SEDs) from the ultraviolet to the far-infrared and characterise their emission-line properties. 
Following a systematic comparison approach, we characterise the properties of the QSO accretion, obscuration, and host galaxies in a sample of $\sim$1800 QSOs at $0.2<z<2.5$, classified into red and control QSOs and matched in redshift and luminosity. 
We  find no  strong  differences  in  the average multiwavelength  SEDs of red and control QSOs, other than the reddening of the accretion disk expected by the colour selection.
Additionally, no clear link can be recognised between the reddening of QSOs and the interstellar medium as well as star formation properties of their host galaxies.
Our modelling of the infrared emission using dusty torus models suggests that the dust distributions and covering factors in red QSOs are strikingly similar to those of the control sample, inferring that the reddening is not related to the torus and orientation effects. Interestingly, we detect a significant excess of infrared emission at rest-frame 2-5 \micron, which shows a direct correlation with optical reddening. To explain its origin, we investigated the presence of outflow signatures in the QSO spectra, discovering a higher incidence of broad [O~{\sc iii}] wings and high C~{\sc iv} velocity shifts (>1000 km s$^{-1}$) in red QSOs as compared to the control sample. We find that red QSOs that exhibit evidence for high-velocity wind components present a stronger signature of the infrared excess, suggesting a causal connection between QSO reddening and the presence of hot dust distributions in QSO winds. We propose that dusty winds at nuclear scales are potentially the physical ingredient responsible for the optical colours in red QSOs, as well as a key parameter for the regulation of accretion material in the nucleus.
}


   \maketitle

\section{Introduction}
Since the discovery of quasars \citep[QSOs,][]{schmidt63}, the most energetic kind of active galactic nuclei (AGN), an in-depth census of QSOs and characterisation of their properties has been carried out. 
These have revealed millions of QSOs \citep[e.g. ][]{flesch19} with a diversity of black hole (BH) masses ($10^{7}-10^{10}$ \Msun) and luminosities ranging up to $\sim 10^{48}$ \ergs, pervading the Universe to its earliest epochs \citep[$z=7.5, $][]{banados18}.
Large optical spectroscopic surveys such as the Sloan Digital Sky Survey \citep[SDSS; e.g.][]{paris18} have played a key role in the identification of QSOs, delivering statistical samples that allow us to achieve a more significant understanding of QSO properties.
Optical spectroscopy is an efficient tracer of QSOs, since their primary emission, the accretion disk, is most luminous in the ultraviolet (UV) and blue optical regime of the spectrum.
The accretion disk emission in QSOs arises from the accretion of the surrounding material onto the BH gravitational potential, heating gas clouds in the immediate surroundings to produce the broad and narrow emission lines characteristic of QSO spectra.
Unobscured QSOs, which constitute the majority of the optically-selected QSO population, are thus characterised by a luminous blue thermal continuum.
However, a small fraction of QSOs exhibit a diversity of redder optical and near-infrared colours. 
Although the origin of these red colours have been the subject of a long list of studies \citep[e.g. ][]{whiting01, wilkes02, urrutia08, glikman07, rose13, kim18, klindt19}, the nature of red QSOs remains an intriguing question in AGN astronomy.

Observationally, several scenarios have been investigated to explain the colours in red QSOs. 
These include red QSOs having an intrinsically red continuum due to particular accretion mechanisms such as higher Eddington ratios \citep[e.g. ][]{richards03, kim18}, or the red colours being just an observational effect due to the luminous contamination from other physical components of the AGN \citep[e.g. synchrotron emission, ][]{serjeant96, whiting01} or the stellar emission from the host galaxy.
The most widely accepted explanation is, however, that the red colours are due to attenuation by dust along the line of sight \citep{webster95,glikman07, klindt19}.
Despite extensive investigations on this topic, the origin of this attenuation and whether the dusty medium is located at nuclear or galaxy scales remains uncertain \citep[e.g.][]{hickox18}.

Generally, obscuration in AGN is mostly linked to the geometrically thick, most likely clumpy, structure, known as the dusty `torus' \citep[e.g. ][]{silva04, fritz06, nenkova08,  hatzi09, mor09,alonsoherrero11, stalevski16}. 
This structure is believed to explain the bulk of the mid- and near-infrared emission, as well as the angle-dependent obscuration of the accretion disk emission and broad line region.
Indeed, the latter leads to the well-established unification model that explains the emission line variations of Type 1 (broad and narrow lines) and Type 2 QSOs (narrow lines only) as an effect of the viewing angle \citep{antonucci93, urry95, ramosalmeida17, hickox18}, which constitutes one of the pillars of AGN astronomy.
Within this framework, a straightforward explanation for the optical colour of red Type 1 QSOs is that it is caused solely by orientation, while red QSOs are intrinsically the same object as their unobscured counterparts.
In this scenario, viewing angles of red QSOs would constitute an intermediate inclination angle between those from unobscured Type 1 QSOs and Type 2 QSOs.

A list of observational evidence, however, suggests that red QSOs present other fundamental differences which cannot be explained by orientation.
Although with a lack of consensus, differences in their host galaxies have been suggested, with host galaxies of red QSOs being more likely to be undergoing mergers (e.g. \citealp{urrutia08, glikman15}; c.f. \citealp{zakamska19}) or have higher star-formation rates than unobscured QSOs (\citealp{georgakakis09}; c.f. \citealp{wethers20}).
\citet{banerji15} reported that the red QSO luminosity functions are flatter than those of unobscured `typical' QSOs, and a higher incidence of low-ionization broad absorption lines (LoBALs) has been found for red QSOs \citep{urrutia09}, indicating the presence of strong outflows.
This set of evidence, although sometimes associated with large uncertainties based on small sample sizes and heterogeneous selections, suggests that red QSOs are a special population, potentially linked to a key phase of galaxy and black-hole evolution.

Indeed, QSOs are expected to play a key role in galaxy evolution as they are invoked as regulators of the growth of their host galaxies through a process known as AGN feedback \citep[e.g. ][and references therein]{harrison17}.
Galaxy evolution models \citep[e.g. ][]{sander88, hopkins08, narayanan10, alexander12} suggest unobscured Type 1 QSOs constitute an evolutionary stage posterior to dusty star-forming galaxies (such as sub-millimetre galaxies). 
Red Type 1 QSOs are often invoked as potential transitioning populations in between these stages.
Although a number of previous observational studies have supported this scenario, statistical samples are required to confirm that these findings are representative \citep[e.g.][]{hickox18}.

Fundamental differences in the radio properties of red QSOs compared to the overall QSO population have recently been discovered \citep{klindt19, rosario20,fawcett20}. 
By matching the comparison samples in luminosity and redshifts, the incidence of radio detection for red QSOs have been found to be up to three times higher than for control QSOs.
Using wide-field radio surveys at high radio frequencies \citep{klindt19, fawcett20} and low radio frequencies \citep{rosario20}, at low and high-resolution, these studies have found that the enhanced radio detection fractions in red QSOs are connected to mostly `radio-quiet' emission (as defined by the radio-to-optical luminosity ratios) and to preferentially compact morphologies.
These observations cannot be explained by the orientation model which would predict the opposite result: an enhancement in compact emission from face-on blue QSOs (due to relativistic beaming) in comparison to red QSOs.
These observations are, consequently, in good support of the evolutionary model whereby red QSOs are an early phase in the evolution of QSOs.

Motivated by the key findings achieved by \gcr{this} comparative approach, in this work we aim to investigate the origin of QSO reddening by studying their photometric and spectroscopic properties across the electromagnetic spectrum.
Indeed, QSOs are prominent examples of multiwavelength phenomena, with characteristic spectral energy distributions (SEDs) from the radio to the gamma rays, which encompass a wealth of information on the physical processes that constitute their nature.
While previous works have studied the SEDs of red QSOs in the optical-near-infrared regime \citep[e.g. ][]{glikman12, lamassa17}, or from the FIR to the UV or X-rays \citep[e.g. ][]{georgakakis09, urrutia12, mehdipour18}, these have focused their studies only on a handful red QSOs (1-13 sources), and without control samples to compare these with.
Building on the robust selection and comparison strategy developed in \citet{klindt19}, we use statistical samples of red and control QSOs to characterise their spectral energy distributions.
In particular, we apply the Bayesian MCMC-based code for SED-modelling \citep[\textsc{AGNfitter};][]{CR16}, which combines multiwavelength-models of host galaxy and AGN physical components specifically tailored for SED studies of QSOs and AGN. 
A clear advantage of \textsc{AGNfitter} over other SED-fitting codes is the inclusion of an independent accretion disk emission component or the `big blue bump', parametrised by a free reddening parameter $\rm E(B-V)_{\rm BBB}$ which is key for the study of red QSOs \citep[e.g.][]{CR16, lamassa17}.

Combining the modelled broad-band photometry with spectroscopic properties inferred from SDSS spectra \citep{rakshit20}, we aim to characterise the nature of red QSOs from a multiwavelength perspective and understand the origin of their red colours. 
In Section \ref{sec:data} we present our selection strategy and the multiwavelength broad-band photometry and spectroscopic catalogues used in this work. 
Additionally, we describe the settings used within the SED-fitting code \textsc{AGNfitter} and other statistical methods used for this analysis.
In Section \ref{sec:seds} we present the SED-fitting results, and combine our measurements with spectroscopic data to infer intrinsic accretion properties of QSOs in \ref{subsec:intrinsicprops}.
In Section \ref{subsec:obscuration} we present our results on physical parameters related to the obscuration in QSOs and in Section \ref{subsec:galaxies} we compare the host galaxy properties inferred from the SED-modelling for red and control QSOs.
Finally, in Section \ref{sec:discussion} we discuss our results and explore the connection of our findings with the incidence of AGN dusty winds in red and control QSOs. 
Throughout this work, we adopt a cosmology with $H_{0}=70$km s$^{-1}$ Mpc$^{-1}$, $\Omega_{\rm m}=0.3$ and $\Omega_{\Lambda}=0.7$.

\section{Data sets and methods}\label{sec:data}

\subsection{Main sample selection}


\begin{figure}
\centering
\centering
    \includegraphics[width=1\linewidth]{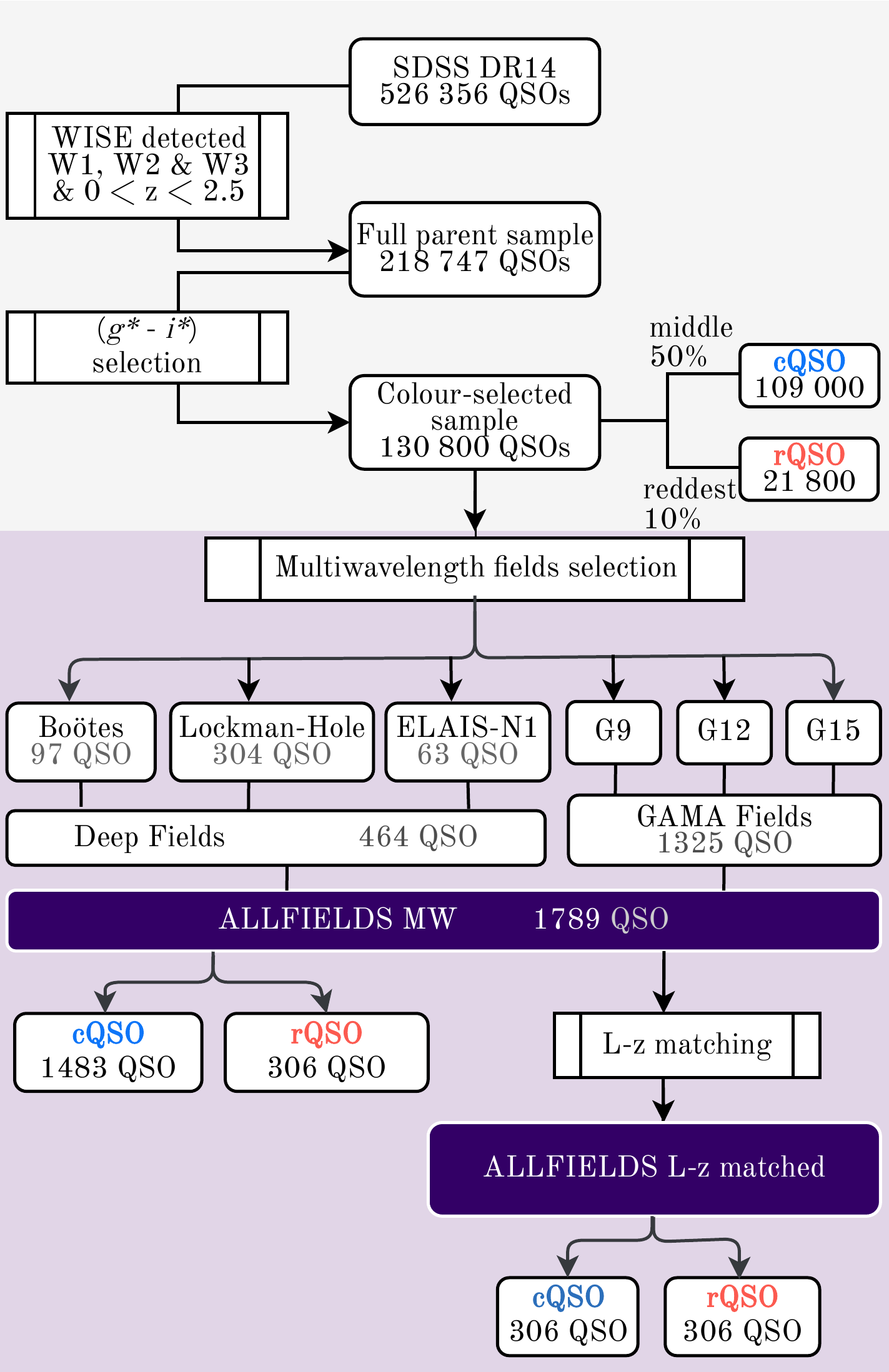}
    \caption{Diagram to summarize our selection process. The shaded area corresponds to the general definition of red and blue QSO samples from the overall SDSS DR14 \citep{paris18}, following the strategy previously presented by \citet{fawcett20, rosario20}. The purple shaded area correspond to the field selection, determined by the availability of deep multiwavelength catalogues provided by the LoTSS data release in three northern fields and the public compilation of the HELP collaboration in the three equatorial fields of the GAMA survey.}
    \label{fig:flowchar}
\end{figure}


The starting sample of our investigation are optically-selected sources defined as QSOs by the SDSS DR14 Quasar catalogue \citep{paris18}.
The SDSS DR14 Quasar catalogue is the largest compilation of QSOs with detailed emission line measurements and black-hole mass constraints \citep{rakshit20} to date.
It consists of 526356 spectroscopically confirmed QSOs (emission line \gcr{full widths at half maximum} FWHM > 500 km\,s$^{-1}$) out to $z=7$, comprising different sets of all archival SDSS spectroscopic campaigns.
These include QSOs spectra selected from SDSS I/II Legacy surveys, targeted QSOs from the Baryon Oscillation Spectroscopic Survey (BOSS) and the extended  Baryon Oscillation Spectroscopic Survey (eBOSS). 
The selection procedure for our main sample is shown graphically in Figure \ref{fig:flowchar} and follows the approach presented by \citet{klindt19, rosario20, fawcett20}.

One key property of the SDSS QSO catalogue is the availability of reliable spectroscopic redshift information for all QSOs in the parent sample.
To provide a robust colour selection in the following steps, the first cut restricts our sample to QSOs at redshifts $z<2.5$, to avoid the Lyman-break feature artificially reddening the optical spectra short-ward of $\lambda \sim 3200 \, \AA$. 
An unbiased multiwavelength comparison of QSOs requires properly accounting for the main fundamental properties that may dominate over other parameters, such as the redshift and the intrinsic luminosities of the QSOs.
A challenge lies, however, in the calculation of the intrinsic bolometric quasar luminosity, which can be very different from the observed luminosity due to dust attenuation. 
Following \citet{klindt19}, we adopt the \gcr{rest-frame} luminosity at 6 $\mu$m ($L_{6 \mu m}$), as the best proxy for the intrinsic luminosity.
The motivation for this wavelength choice relies on the rest-frame MIR emission arising from the dust structure heated by the accretion disk, and not suffering any further extinction.
The 6 $\mu$m luminosities $\Lsixum$ are calculated for each quasar based on the extrapolation of near- and mid-IR measurements from the WISE All Sky survey \citep{wright10}.
To perform this calculation consistently for all QSOs we thus require a second cut, selecting only sources with WISE detections at SNR$>3$ (W1-W3).
The implications of this selection have been investigated by \citet{rosario20}, showing this cut only focuses our study on the most powerful AGN, without any significant bias on colour.
Recently, a study of quasar bolometric luminosities from $z=0-6$ \citep{shen20} indeed shows that the infrared luminosity is ideal for such a study as the required correction factor from the monochromatic to bolometric luminosities remains almost constant across the whole redshift range, in particular for QSOs of luminosities $\rm \log L_{\rm bol}>44$ \ergs.
The number of optically (SDSS) selected QSOs at $z<2.5$, with WISE detections in at least three bands is 218747 in total. 


The SDSS QSO catalogues also provides photometric information from the SDSS \textit{ugriz} bands.
Based on the $g^{*}$\,($4770\,\AA$) and $i^{*}$\,($7625\, \AA$) band photometry, our \gcr{third} selection criterion uses the \textit{g$^{*}-$i$^{*}$} colours, to define the populations of red QSOs and a control sample of `normal' QSOs, following the same strategy presented by \citet{rosario20} and \citet{fawcett20}. 
We define red QSOs as the 10 per cent of the total QSO population with reddest \textit{g*$-$i*} colours, and control QSOs as the 50 per cent of the total QSO population closest to the median value of the total population colours.
In particular, to account for the redshift dependence of a colour selection, the red and control samples were defined by binning the \textit{g$^{*}-$i$^{*}$} distribution as a function of redshift, in contiguous redshift bins of 1000 sources each.
Following this classification scheme, from the  218747 optically (SDSS) selected QSOs at $z<2.5$, 21 800 are classified as red QSOs, 109000 are classified as control QSOs.

While the entire QSO catalogue covers a region of 9376 deg$^{2}$, for the purposes of this investigation we select QSOs located within six survey fields which cover $\sim 175 \deg^2$ in total and include three surveys in the northern sky: the Bo\"otes field, the Lockman-Hole Field, and the ELAIS-N1 Field, and three areas in the equatorial sky covered by the GAMA surveys \citep{driver11}: G9, G12, and G15.
These fields were selected for two reasons. First, they have rich multiwavelength data from multiple surveys from the FIR to the UV, including deep Herschel coverage, some of which were recently compiled by the Herschel Extragalactic Legacy Program (HELP) (see more details in Section \ref{subsec:photometry}), and second, all these fields have deep LOFAR surveys, including the LoTSS Deep Fields surveys \citep{tasse20, sabater20} and the LOFAR GAMA surveys (Williams, Hardcastle et al. \textit{in prep.}).
The total number of selected QSOs defined as red or control QSOs in each field are summarised in Figure \ref{fig:flowchar}, comprising 1483 control QSOs and 306 red QSOs across all six fields, making a total of 1789 sources included in this study.

\subsection{Lz-matched subsample selection}\label{subsec:Lzselection}

\begin{figure}
    \centering
    \includegraphics[trim={ 0.3cm 0.2cm 0.5cm 0cm},clip,width=\linewidth]{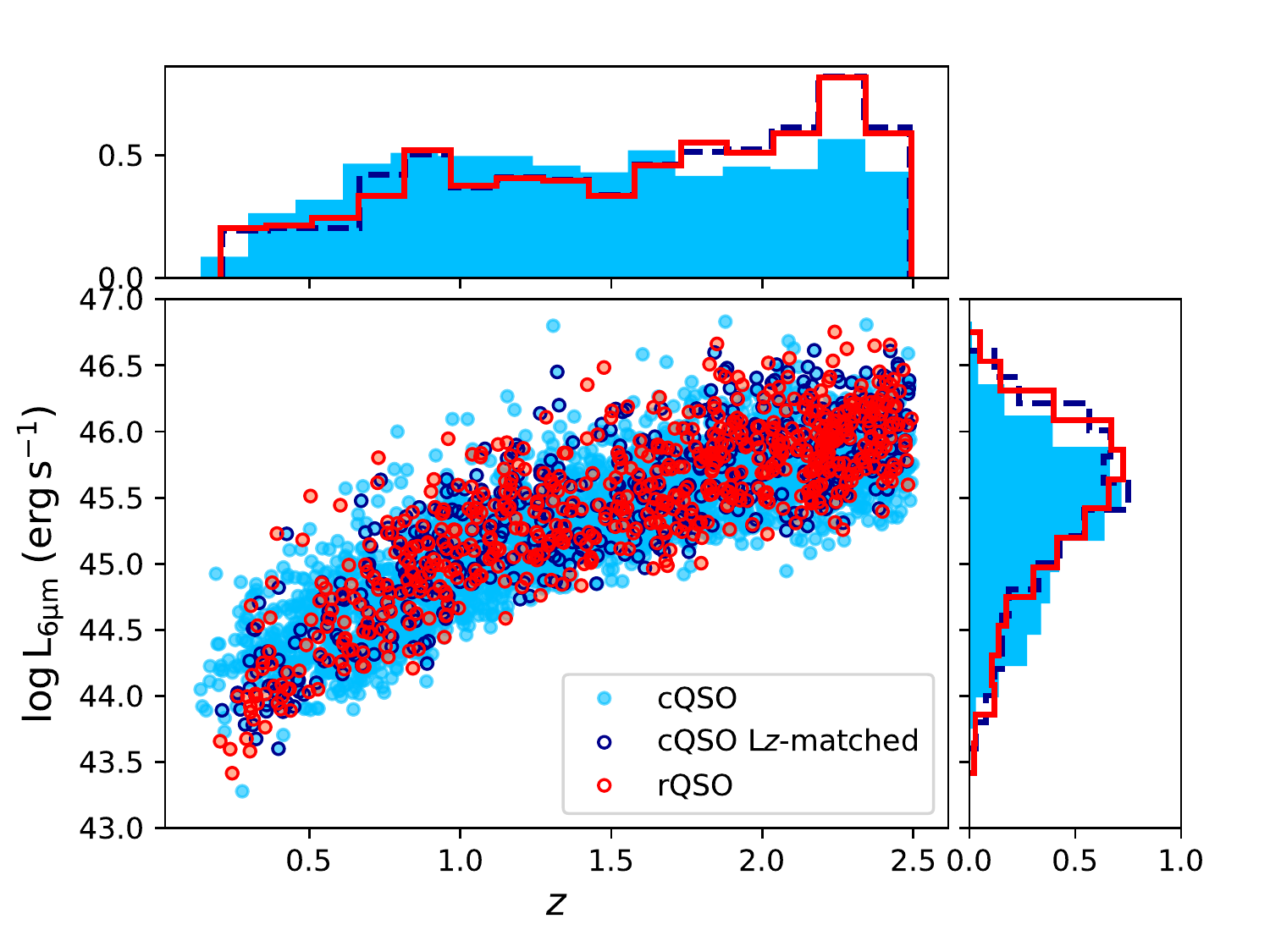}
    \caption{Monochromatic  6 \micron\, luminosities (L$_{6\mu \rm m}$) of red QSOs (red filled circles), control QSOs (sky-blue filled circles) and Lz-matched control QSOs (dark-blue circles) as a function of redshift are plotted in the central panel. Side panels show normalised histograms of the L$_{6\mu \rm m}$ and redshift distributions for each population, shown as solid lines (red QSO), dashed lines (L-z matched samples) and shaded areas (entire control sample) of respective colours. By comparing the distributions of the red QSOs and the Lz-matched control QSO sample, it can be clearly seen that equivalent QSO properties are being sampled.  }
    \label{fig:L6z}
\end{figure}

In Figure \ref{fig:L6z} we show the redshift and L$_{ 6 \mu \rm m}$ distributions of red and control QSOs that result from the sample selection described above.
In the central panel red QSOs are represented by red circles, and as red dashed histograms in the side panels.
The control QSOs are shown as sky-blue circles and sky-blue filled histograms in the side panels.
The single parameter histograms in the side panels show that our selection has produced samples of slightly different redshift and $\Lsixum$ distributions.
In particular, the difference is significant at redshifts $z>1.5$, where red QSOs appear to peak, and in connection with this, a larger fraction of the selected red QSOs have 6 $\mu \rm  m$ luminosities larger than $\log \Lsixum = 45.5$.

In order to overcome these disparities and guarantee comparable quasar properties between red and control samples, that is to day equivalent distributions of luminosities and redshift, 
we assume a conservative approach and define a sub-sample of control QSOs that is directly matched in redshift and luminosity to our selected red QSO sample. 
Given the larger number of control QSOs compared to red QSOs (5 times larger), we select the closest control QSOs in $\Lsixum$ and $z$ for each red quasar, matched using the Nearest Neighbour method defining a two-dimensional metric on the  $\Lsixum-z$ space. 
Using this approach we define the $\Lsixum-z$-matched sample, which contain 612 sources, composed of the 306 red QSO, and 306 L-z matched control QSOs. 
In Figure \ref{fig:L6z} we overplot the $\Lsixum-z$-matched subsample as dark blue circles in the central plot and as a solid dark blue histogram in both side panels.
The similarity achieved in the distributions of red and Lz-matched control QSOs is now compelling, providing a sub-sample for our further comparative investigation between red and control QSOs.

\gcr{We note that the entire analysis included in this paper was performed using both the complete and the $\Lsixum-z$-matched samples, finding that our main results do not change irrespectively of the sample choice (in agreement to what was previously found in the radio properties by \citealp{klindt19}).
Thus, for the purpose of clarity, throughout this paper we use `control QSOs' or `cQSOs' and dark blue markers to refer to the control sample matched in $z$ and $L_{6\mu m}$ to the sample of red QSOs (or `rQSOs'), which will be shown as red markers.}

\subsection{Multiwavelength photometry}\label{subsec:photometry}

For consistency with the sample definition and colour selection, we constructed the multiwavelength SEDs based on the SDSS \textit{ugriz} bands for the optical regime, and the four WISE bands in the infrared \citep{paris18}.
For the three northern Deep Fields we complement these basic SEDs with the aperture-matched photometry compiled for the LoTSS data release \citep{kondapally20}.
For the equatorial GAMA fields we complement this with aperture-matched photometry from the Herschel Extragalactic Legacy Project HELP project \citep{HELP19}, utilising their merged catalogues dmu32\footnote{
1. http://hedam.lam.fr/HELP/dataproducts/dmu32/dmu32_GAMA-09/\\
2. http://hedam.lam.fr/HELP/dataproducts/dmu32/dmu32_GAMA-12/\\
3. http://hedam.lam.fr/HELP/dataproducts/dmu32/dmu32_GAMA-15/} to construct the respective multiwavelength SEDs. 

\gcr{We use optical position coordinates to find the multiwavelength counterparts to the SDSS QSOs in our sample, using a search radius of 1 arcsec with respect to the SDSS positions.}
Given the point-source nature and large luminosities of the optical QSOs, this process is straightforward and overall we find counterparts for 99.9 \% of our sample.
In the three northern fields, 97\% of the covered sources (469) have counterparts in the multiwavelength bands. For the remaining sources in these three fields (17) we construct the SEDs based on SDSS and WISE data only. \gcr{We note that no biases exist towards red or control QSOs for these 17 sources.} 
Given the small number of those without counterparts in the LoTSS catalogue, the impact of the inclusion of these sources on our results is negligible.
In the three GAMA fields, we find catalogued counterparts for all QSOs from our main sample.

In the three northern fields, we complement the SDSS photometry with the aperture-matched UV-optical and infrared data available in each field, which includes FUV and NUV from GALEX surveys, and a wide suite of ground-based optical and near-infrared filters reaching out to the K-band. 
For a detailed description of the photometry in each of the three fields please refer to \citet{kondapally20}.
In the equatorial fields, the HELP catalogues compile calibrated data from a large number of surveys, including the GAMA survey data\footnote{In the GAMA fields we exclude all DECAM bands as these cover only a small fraction of the sky.} in the optical and the NUV and/or FUV GALEX measurements when available.
In the near-infrared regime, these are complemented with Y, J, H, Ks bands from both the VISTA and UKIDSS survey.

To cover the MIR regime, the WISE bands are complemented by four IRAC bands at 3.6, 4.5, 5.8 and 8.0 $\mu$m and the 24 $\mu$m MIPS band in the three northern fields.
Since the Spitzer IRAC and MIPS bands are not available in the equatorial fields, here we only use the WISE data.
Given that our sample selection already requires strong detections for all our sources in at least three WISE bands, and the IRAC and WISE bands cover similar wavelength regimes, the different sensitivities and lack of IRAC coverage in these field do not significantly affect our conclusions. 

Finally, the infrared SEDs of the QSOs in our selection are well sampled by measurements in the 100, 160, 250, 350, 500 $\mu$m bands from deep Herschel surveys, including the HerMES project \citep{oliver12} for the three northern fields and the H-ATLAS project \citep[e.g.][]{valiante16} for the GAMA fields.
In the northern fields, these have been probabilistically deblended using the XID+ tool \citep{hurley17} as presented by \textit{McCheyne et al. in prep}.
In a similar manner, in the equatorial fields HELP computes the deblended Herschel emission for all sources in their catalogues based on optical and MIR priors using the same XID+ method, recovering deblended fluxes in the Herschel bands.
While the Herschel flux measurements were deconvolved and robustly assessed, a significant fraction of these are measurements with large uncertainties and in a few cases consistent with zero. 
In the three northern fields, which were observed with the Herschel PACS and SPIRE instruments with up to 10 times longer exposures per square degree than the H-ATLAS fields \citep[e.g. ][]{lutz14}, 96.6 \% have robust Herschel measurements with uncertainties lower than 70\%, inconsistent with zero.
In the three equatorial GAMA fields, which are part of the H-ATLAS survey, 87 \% of the sources have Herschel estimates.
From these, a significant fraction of sources have \gcr{large photometric uncertainties and consequently fluxes consistent with zero }: $\sim 23\%$ of the total sample in the case of the SPIRE 250 \micron\,band, and up to $\sim 59\%$ in the case of the SPIRE 500 \micron.
In order to properly account for this probabilistic information, and robustly infer star formation properties from such diverse statistical properties a Bayesian method is imperative, and we expand on this in Section \ref{subsec:agnfitter}.

In total the SEDs for this study consist of 25-30 photometric data points, from the FIR to the UV.
\gcr{We note that we can reliably discard a possible contribution from synchrotron emission to the total SEDs based on available deep low-frequency radio imaging of these fields \citep[LoTSS, ][]{tasse20, sabater20, kondapally20}.  
We expand on the radio properties of our sample in a companion paper (Calistro-Rivera, \textit{in prep}).
To summarise the results relevant to this paper, we find that most QSOs are in the radio quiet regime, in line with the findings by \citet{klindt19,fawcett20, rosario20}. 
In particular, we find the synchrotron emission contribution to the FIR-to-UV SED is negligible for the large majority (85\%) of the sources in our sample, relatively minor for 11\% of the sources (contribution to the infrared luminosities < 10\%), and is significant only for the remaining 4\%.
}

\subsection{Spectroscopic properties}

In addition to the photometry, we compiled spectroscopic information for our QSO sample.
The spectra of all SDSS DR14 QSOs were analysed, decomposed and modelled by \citet{rakshit20}, providing a catalogue of spectral properties which we use for this study.
The estimated spectral properties in this catalogue include line fluxes, FWHM estimates for broad and narrow components of several emission lines.
In particular in this work we use broad emission line properties from the H$\beta$, Mg~{\sc II}, and C~{\sc IV} lines, including their FWHMs estimates and associated uncertainties.
\gcr{
In \citet{rakshit20}, the H$\beta$ and Mg~{\sc ii} lines were modelled with broad and narrow components, where multiple Gaussians were used to fit the broad component, and a single Gaussian was used to fit the narrow component. 
During the fitting, the narrow component was allowed to have a maximum FWHM of 900 km s$^{-1}$ while each of the Gaussians in the broad component had FWHM >900 km s$^{-1}$.  
While the spectral quantities reported for H$\beta$ and Mg~{\sc ii} were estimated from the broad component, in the case of C~{\sc iv} the total profile was used without the subtraction of a narrow component, because of the ambiguity in the presence of narrow components in these lines \citep{shen11, shen20}.
A detailed description of the spectral fitting can be found in \citet{rakshit20}.}

Additionally, we also use the spectral fitting output by \cite{rakshit20} on the [O~{\sc iii}]$\lambda$5007 emission line (through private communication), covered by the SDSS spectral band only for sources with $z<1.$
A double Gaussian model was used to represent the [O~{\sc iii}] emission line; one for the core narrow component and another for the wing component, to characterise potential outflows. 
More details are discussed in Section \ref{sec:discussion}.

Finally, we inspect continuum measurements reported by \citet{rakshit20} for comparison to our own estimates from SED fitting, such as bolometric luminosities and monochromatic luminosities.
For control QSOs we find a general agreement, as expected.
Since the work by \citet{rakshit20} does not include corrections for QSO reddening, the reported bolometric luminosities are underestimated for red QSOs when compared to the reddening-corrected continuum results.
In Section \ref{subsec:intrinsicprops} we discuss these differences and use the reported emission line properties together with reddening-corrected continuum information from our SED-fitting analysis, to estimate virial $M_{BH}$ and Eddington ratios for control and red QSOs.

\begin{figure*}
    \centering
    \includegraphics[trim={0 0 0 0},clip, width=0.49\linewidth]{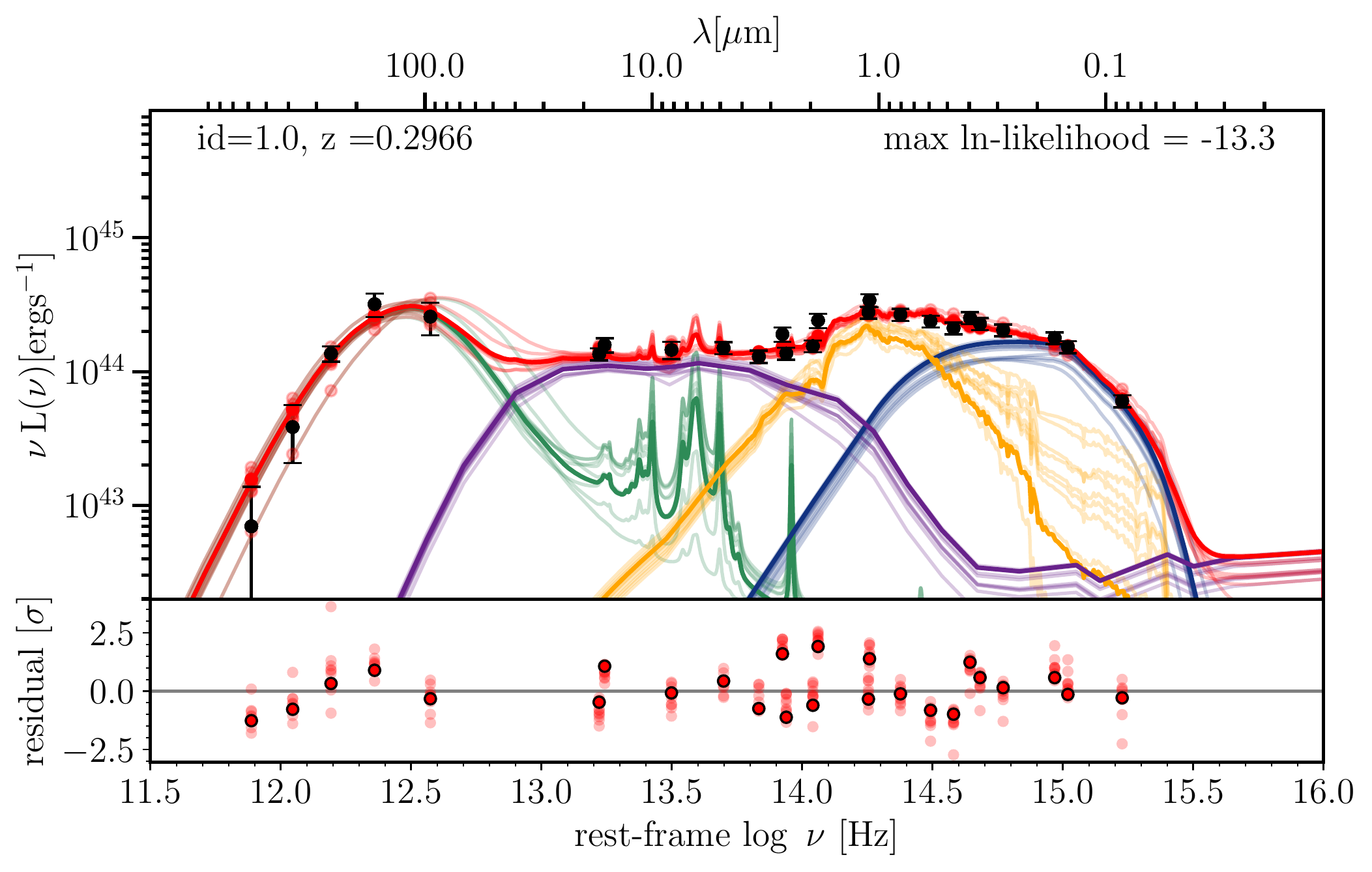}
    \includegraphics[trim={ 0 0 0 0 },clip, width=0.48\linewidth]{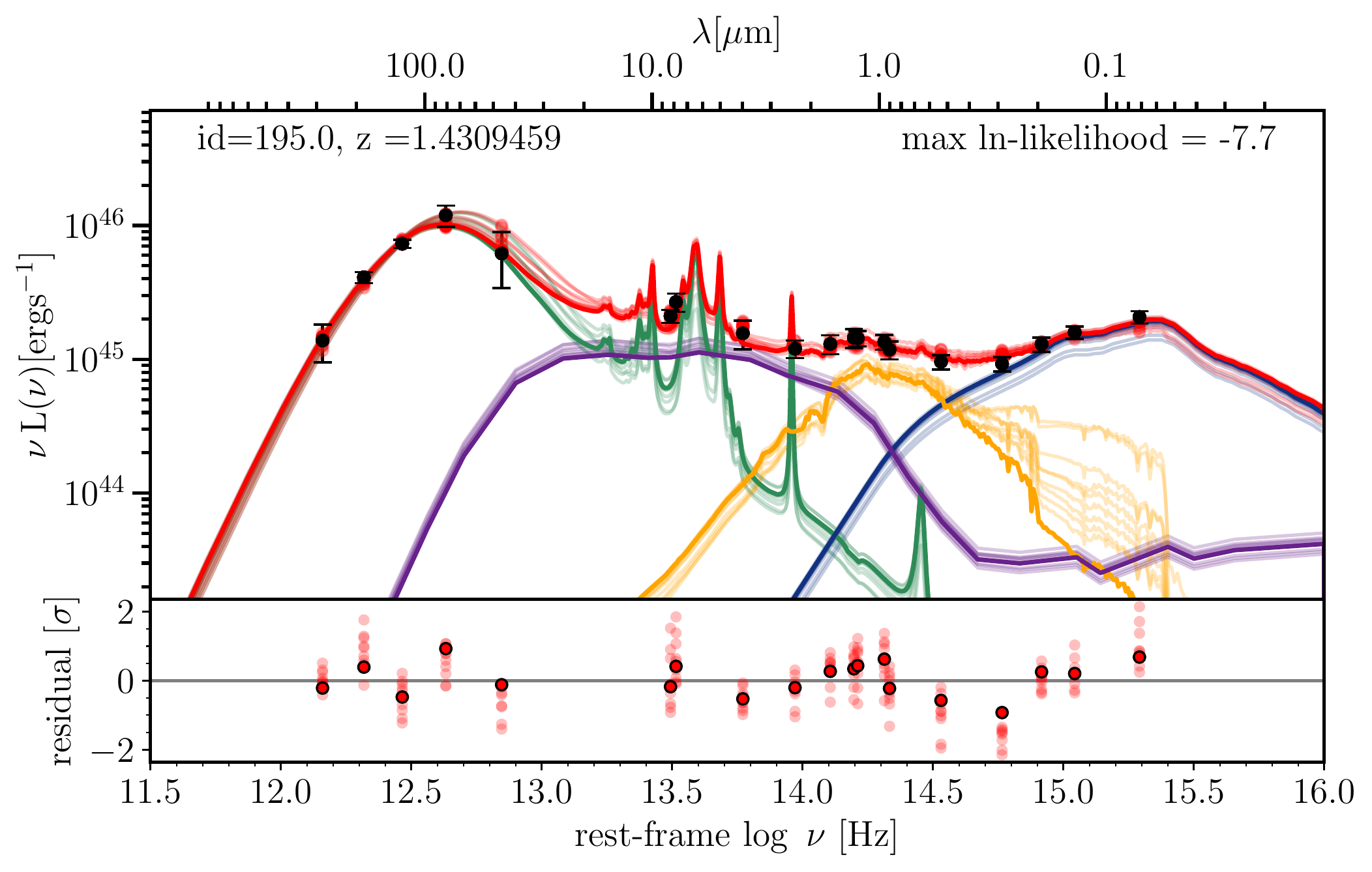}
    \includegraphics[trim={ 0 0 0 0 },clip, width=0.49\linewidth]{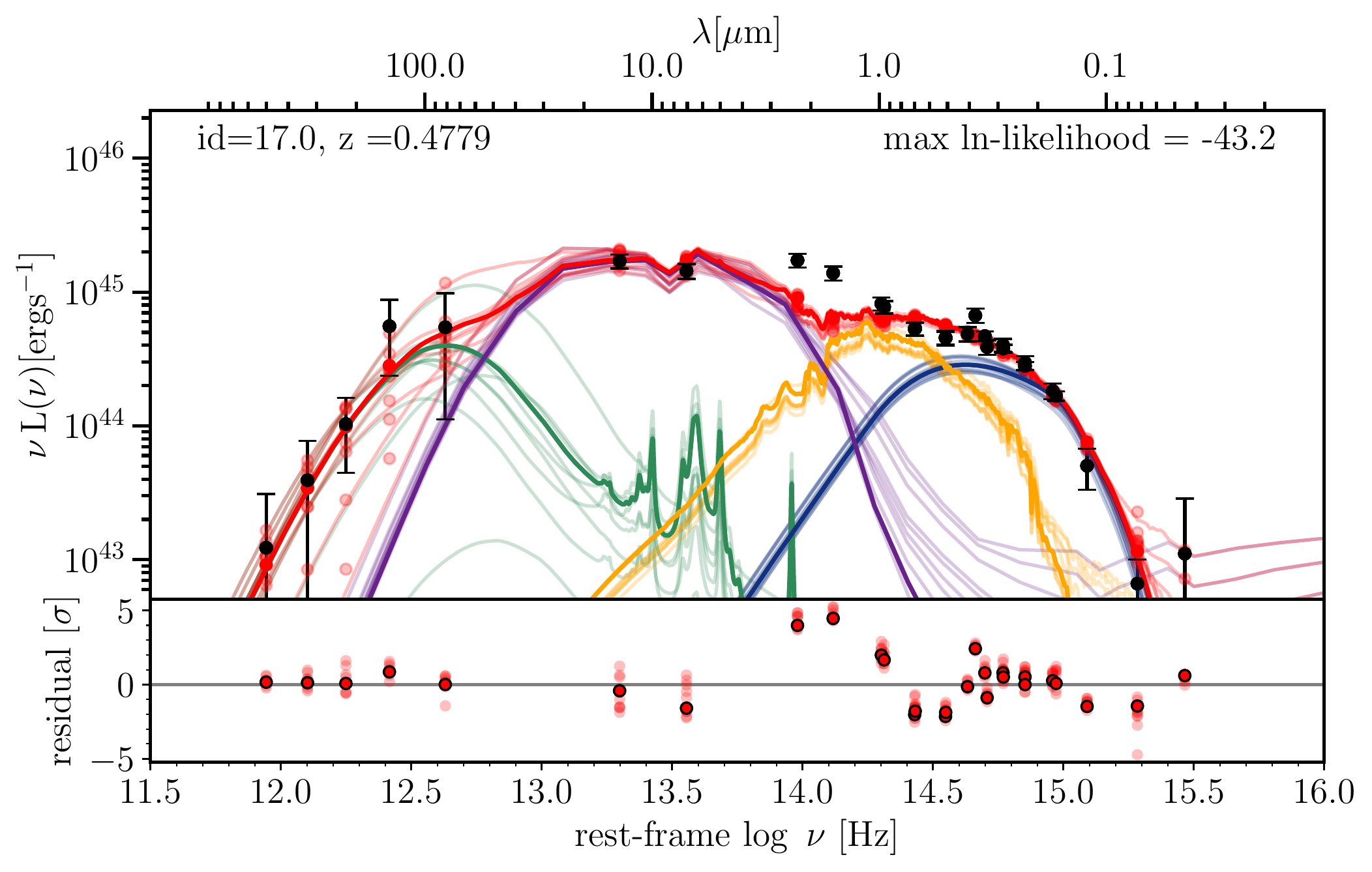}
    \includegraphics[trim={ 0 0 0 0 },clip, width=0.48\linewidth]{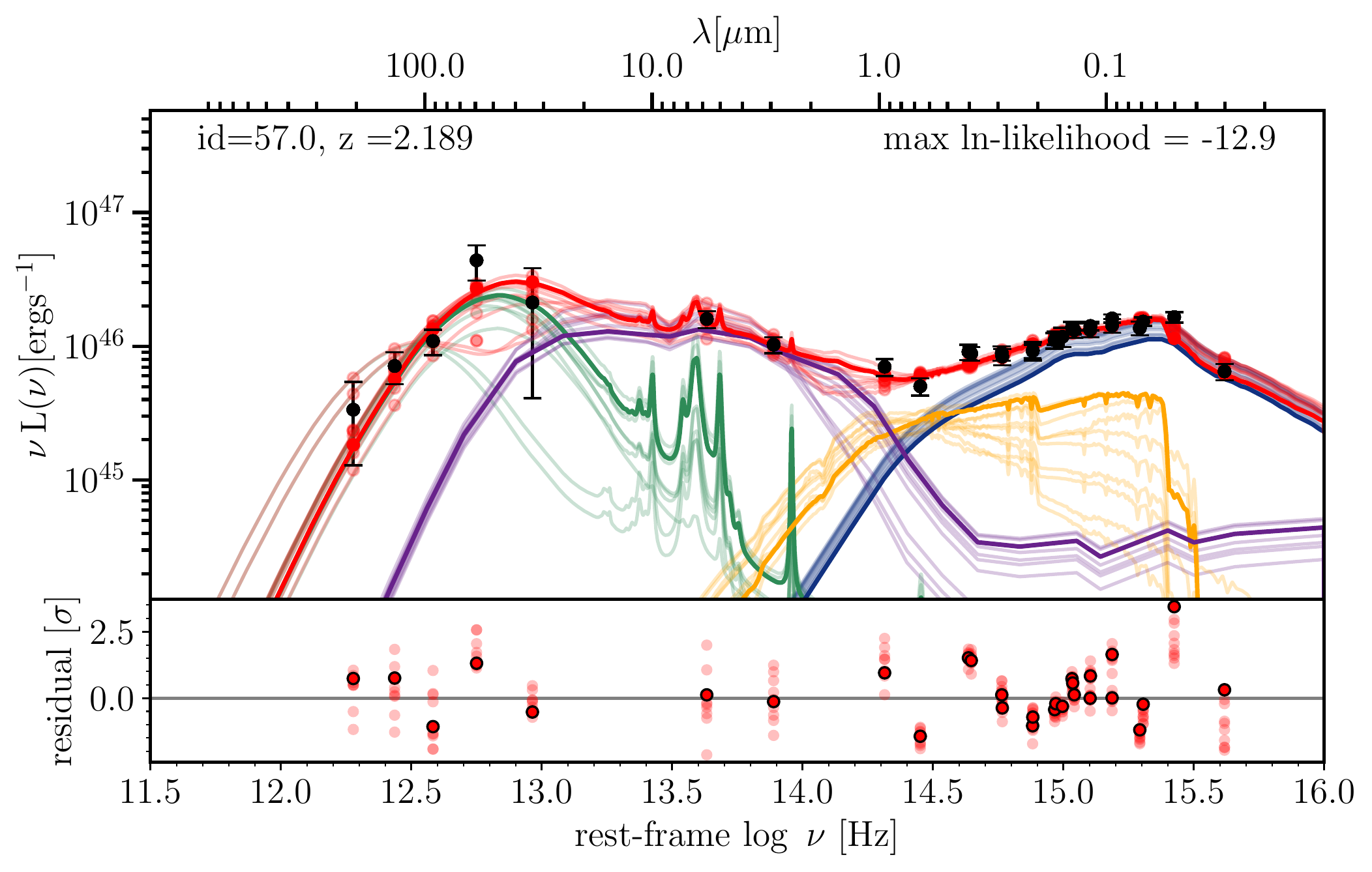}
     \includegraphics[trim={ 0 0 0 0 },clip, width=0.49\linewidth]{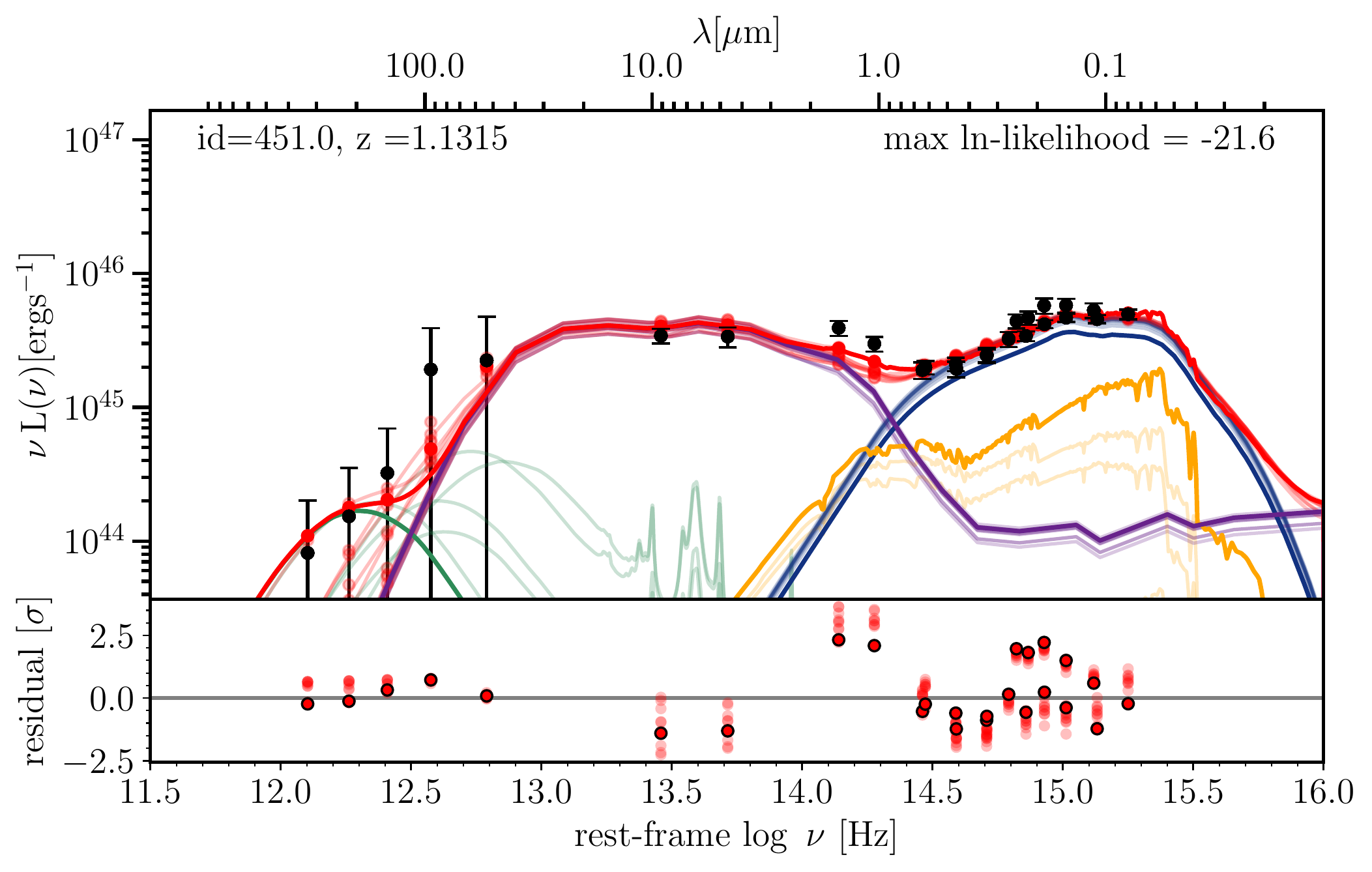}
     \includegraphics[trim={ 0 0 0 0 },clip, width=0.48\linewidth]{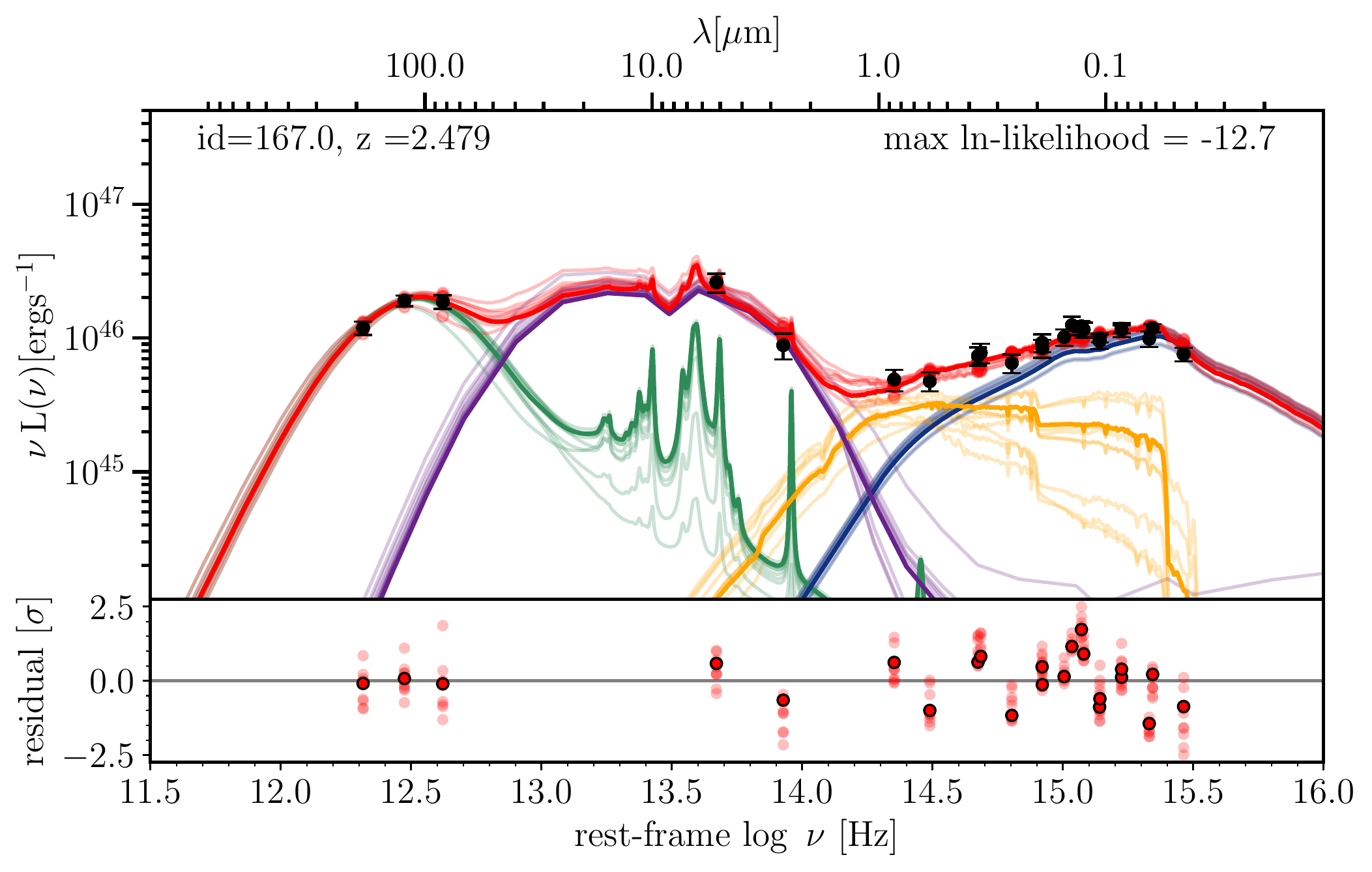}
    \caption{Examples of the SED decomposition with AGNfitter. In the main panels the photometric data points are shown with black error bars. The best fit and ten realisations of the posterior distributions of the fit are drawn as coloured solid, and transparent lines, respectively. The realisations display the uncertainties associated with the fits. The different colours represent the total SED (red lines), the accretion disk emission (blue lines), the host galaxy emission (yellow lines), the torus emission (purple lines) and the galactic cold dust emission (green lines). \gcr{The lower panels show the residuals expressed in terms of significance given the data noise level for the best fit and the 10 realisations.}
    Through this selection of SED fits we exemplify the diversity of the obscuration levels found. 
    Two examples of red QSO SEDs which clearly present an infrared excess emission at $\sim 3 \, \mu$m, as discussed in Section \ref{subsubsec:residuals}, are included in the two lower left panels.
  }
    \label{fig:SEDs}
\end{figure*}
\subsection{Fitting spectral energy distributions with AGNfitter}\label{subsec:agnfitter}

To infer physical properties of the active nuclei and their host galaxies we model the panchromatic spectral energy distributions constructed in Section \ref{subsec:photometry} using an updated version of the SED-fitting code \textsc{AGNfitter} \citep{CR16}. 
The total AGN SED model in \textsc{AGNfitter} consists of four physical components: the QSO accretion disk, the dusty torus, the stellar populations, and the galaxy cold dust component.
The accretion disk emission model consists of a modified version of the empirical template by \citet{richards06} attenuated by a free reddening parameter E(B-V) which is applied following the Small Magellanic Cloud reddening law by \citet{prevot84}.
The dusty torus emission is modelled using the model by \citet{silva04}, which includes a library of 80 torus templates parametrised by column density in the range of $\rm 10^{21}\,\rm cm^{-2}< N_{\rm H}< 10^{25}\, cm^{-2}$, which is associated with the inclination angle.
While the fitting machinery and models employed for the inclusion of the host galaxy contribution are mostly the same as those described by \citet{CR16}, we made a few changes to the publicly available version which we describe below.

The new version of \textsc{AGNfitter} (\textsc{agnfitter2.0}, in prep.) was used which increases the flexibility for the easy inclusion of new models. 
One new model included in this version is the cold dust SED library developed by \citet{schreiber18}.
These models were constructed using the dust continuum models from \citet{galliano11} and have been calibrated to include high redshift ($z=2-4$) galaxies.
This model depends on three free parameters: the dust mass ($\rm M_{\rm dust}$), the dust temperature ($\rm T_{\rm dust}$), and the mid-to-total infrared colour (IR8 = LIR/L8), which measures the relative contribution of polycyclic aromatic hydrocarbon (PAH) molecules to the total infrared luminosity.
The main advantage of these models compared to previously used models lies especially in the MIR regimes, where the \citet{schreiber18} library allows for higher flexibility in the PAH contributions, which is necessary for star forming galaxies at higher redshift.
A second advantage of these models lies in the possibility of exploring a different parameter space than our previous low-z calibrated models, where the dust temperature is of high relevance as it has been observed to increase at higher redshifts \citep{magnelli14, bethermin15}.
A second variation in the model libraries is the inclusion of a larger and more detailed template library of the stellar population synthesis model.
\gcr{For this we use \citet{bruzual03}, assuming a \citet{chabrier03} initial mass function and an exponentially declining star-formation history parametrised by the timescale $\tau$ as $\psi(\mathrm{age}) \propto e^{-\mathrm{age}/ \tau}$. The \citet{bruzual03} model library in \textsc{AGNfitter} has now been extended  to include metallicity as a free parameter, in contrast to the previous version of the code where our library assumed a solar metallicity.} More details on this model can be found in \citet{CR16}.

The new version of \textsc{AGNfitter} also includes the addition of three informative priors, specifically tailored for this QSO population.
A prior on the maximum expected contribution of the host galaxy at rest-frame UV has been updated, based on the expected unobscured UV galaxy luminosity functions presented by \citet{parsa16}, which are calibrated using data from low to high redshift.
This prior is applied on the ratio of accretion disk to galaxy contribution ratios at 1500 $\AA$, in order to avoid unphysically large stellar masses, giving preference to a higher AGN contribution when the rest-frame data at 1500 $\AA$ is 10 times higher than the maximum expected by the \citet{parsa16} galaxy luminosity function at the given redshift.
To achieve a conservative and smooth prior we use a wide Gaussian function, which applies this condition unless the data strongly prefers the galaxy contribution.

The probability that the accretion disk emission in QSOs outshines the host galaxy contribution in the optical is large, thus limiting the inference of the stellar mass parameters in the case of the most powerful QSOs.
Due to this we observed that the fitting of a fraction of the QSO SEDs in our sample resulted in stellar masses that are unrealistically low for the host galaxies of such powerful SMBHs.
In order to use information on a lower limit of stellar masses in a conservative manner, we applied a prior of a wide Gaussian shape on the normalization parameter of the host galaxy (\textsc{GA}) with a mean value of \textsc{GA}$=4.5$ and a standard deviation of $\sigma=1.5$.
We choose these values by taking into account that the normalization parameter \textsc{GA} correlates strongly with the output on total host-galaxy stellar-masses, where \textsc{GA}$<3$ corresponds approximately to total stellar masses of $\rm M_{*}<10^{9}$\Msun, an unlikely scenario for our sample, with a scatter that depends on the variety of SED shapes. 
We tested the robustness of this prior finding clearly different prior and posterior distributions for the parameters. This shows that while supporting the estimation of physical stellar masses, the prior does not entirely determine the posterior probability distribution but there is significant constraining power from the data. 

Finally we include a conservative prior to consider an \textit{energy balance} between the dust-attenuated emission from the stellar populations and the reprocessed cold dust emission.
In particular, we assume that the energy budget emitted by the cold dust in the infrared is \textit{at least} as high as the energy comprised by the attenuated stellar emission.
In contrast to other methods used in the literature, we adopt this conservative approach, so that our model is flexible enough to account for uncertainties evident in dusty star forming galaxies; in these sources, large uncertainties on light to mass ratios, accuracy of dust attenuation laws, and the increasing number of observed uncorrelated distributions of dust and stars, may challenge more stringent energy-balance assumptions \citep[e.g.][]{CR18, buat19}. 

\gcr{We apply the SED-fitting routine on the entire sample of 1789 QSO SEDs, achieving high quality fits (ln_like<50) for  92\% of the sources, and fair quality fits for another 5\%. 
Although we fit the entire sample, as stated in Section \ref{subsec:Lzselection} we consider only the red and control QSO samples matched in redshift and luminosity for the analysis and interpretation of the results in Sections \ref{sec:results} and \ref{sec:discussion}}.
Examples of the fitting results can be found in Figure \ref{fig:SEDs}, for a few cases of control and red QSOs.
We note that an important assumption in the physical modelling in \textsc{AGNfitter} is that any variations in the optical-UV shapes of the accretion disk emission SED are explained by reddening through dust attenuation. 
This assumption would be wrong if the red colours would be due to intrinsic accretion properties instead. 
\gcr{However, extensive evidence exists in the literature \citep[e.g.][]{richards03, glikman12, glikman13, kim18, wethers18, klindt19, temple20} which supports the dust-reddening scenario as the most plausible description of the data.
In particular, the QSO reddening model in \textsc{AGNfitter} follows a Small Magellanic Cloud (SMC) attenuation law, which is assumed to be a good description for the reddening in Type 1 SDSS QSO spectra \citep[e.g.][]{glikman12}.
In addition to the mentioned studies,  in a forthcoming investigation (Fawcett et al. \textit{in prep}) we apply detailed spectral fitting using different reddening models to X-Shooter UV--near-IR spectroscopy of a sample selected in the same manner as this study, demonstrating that the dust-reddening scenario is a robust description for our QSO sample.
Finally, using SDSS spectroscopic data and reddening-corrected fluxes, we find no significant differences in the intrinsic properties of the red and control QSOs (see Section \ref{subsec:intrinsicprops} for more details), validating our dust-attenuation approach.}

\begin{figure*}
    \centering
    \includegraphics[trim={ 0.2cm 0.2cm 0.2cm 0cm},clip, width=0.9\linewidth]{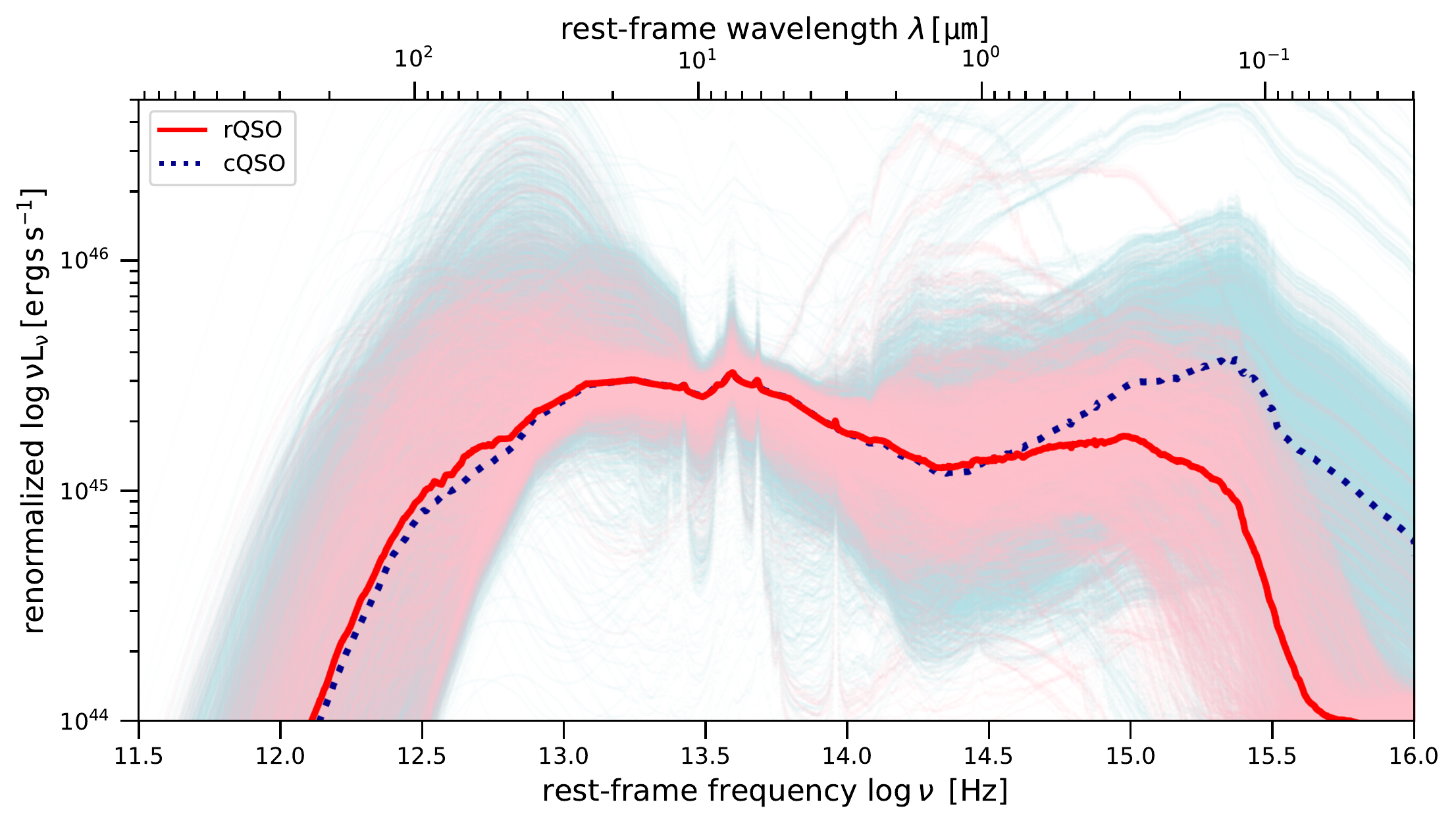}
    \caption{ \gcr{Ten modelled SEDs are plotted for each source (reconstructed from 10 randomly-picked realisations of the posterior PDFs) as transparent red (red QSOs) and blue lines (control QSOs) to display the uncertainties. Solid red and dotted blue lines depict the composite median SED  for the red and control QSO population respectively, matched in redshift and $\Lsixum$.} At frequencies $\log \nu <14.5$ Hz, strikingly similar composite SEDs can be observed for the red and control QSO populations. At frequencies $\log \nu >14.5$ Hz a clear difference in the composite SEDs can be observed which is the signature of dust reddening consistent with the optical colour selection.}
    \label{fig:seds_median}
\end{figure*}

\subsection{Statistical comparative methods}\label{subsec:stats}
The Bayesian method inherent to the \textsc{AGNfitter} code allows us to sample the probability density functions for all of the output physical parameters.
The \textsc{AGNfitter} output therefore includes the parameter posterior PDFs for each source sampled using 100 random realisations.
To achieve robust comparisons of the properties of red and control QSOs in this study, we need to account for the uncertainties and shapes of the posterior PDFs of each source.
To this end we produce composite probability distributions for the red QSO and control QSO populations, as a superposition of the PDFs for the single sources within each group. 

For purposes of visualisation, we apply a Gaussian kernel-density estimator (KDE) with a given bandwidth (specified in each plot).
To assess the significance of the differences between the distributions we apply a bootstrap version of the non-parametric two-sample Kolmogorov–Smirnov (KS) test.
\gcr{This method consists in defining bootstrapped pairs of red and control QSO samples and calculating the KS probability through the application of the two-sample KS test. 
We note that, throughout Sections \ref{sec:results} and \ref{sec:discussion}, we consistently use red and control samples matched in redshift and luminosity for the comparisons, if not otherwise stated.}
To quantify the distribution of KS p-values that arise from the whole composite distributions, we perform 3000 bootstrap iterations, after testing for convergence of the results at this iteration number.
Finally, the median and percentiles values of the KS p-values represent the probability that the parameter distributions for the two samples are drawn for the same distribution, i.e. the distributions have no significant differences.
In particular, median KS p-values of $p<0.05$ would imply that we find statistically significant differences in properties of red and control QSOs (Sections \ref{sec:results} and \ref{sec:discussion}).

\section{Results}\label{sec:results}

\subsection{The far-infrared--UV SEDs of blue and red QSOs} \label{sec:seds}
\begin{figure*}
    \centering
    \includegraphics[trim={ 0.25cm 0.2cm 0.5cm 0cm},clip, width=
    0.85\linewidth]{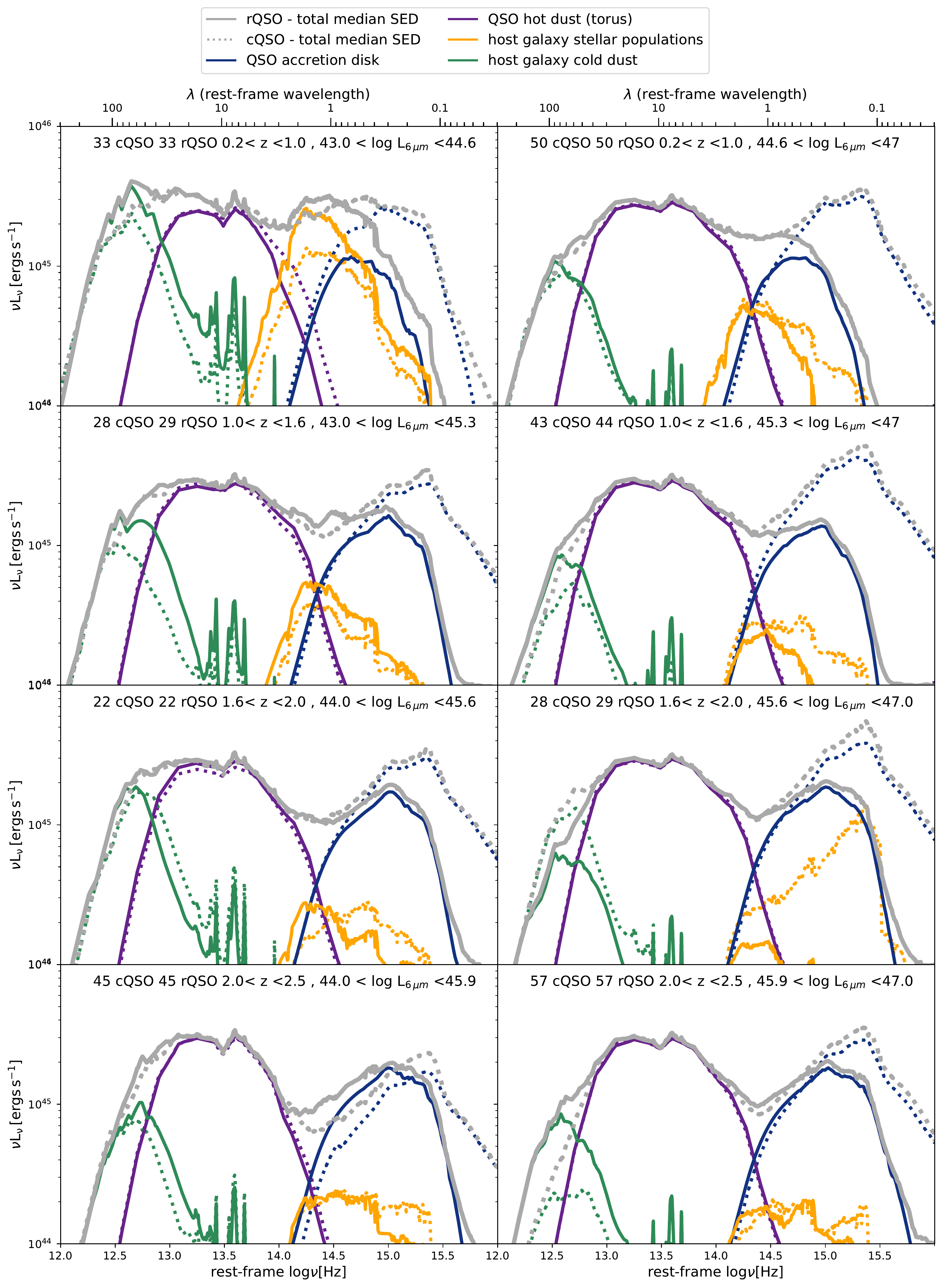}
    \caption{ \gcr{Median composite SEDs  are shown for the red QSO sample (solid lines of different colours) and matched control QSO samples (dotted lines of different colours), now split into different bins of L$_{6 \mu m}$ (left-right) and $z$ (top-bottom). All composite SEDs have been renormalised to the same L$_{6 \mu m}$ for purposes of visualisation.} The grey solid and dotted lines represent the total SEDs for the red and control samples, respectively. The remaining colours codify the different physical components as described in the legend of Figure \ref{fig:SEDs}.  The luminosity and redshift bins considered and the number of sources included in each bin are annotated. The contribution of the stellar emission of the host galaxies has a clear contribution to the total SEDs in the lower redshift bins and to luminosities of L$_{6 \mu \rm m}<45$.   A remarkable similarity in the composite torus SEDs  of red and control QSOs across all redshift bins can be seen.}
    \label{fig:seds_Lz_descomp}
\end{figure*}
We show average SEDs for the different QSO populations in Figure \ref{fig:seds_median}. All have been renormalised to have the same $\Lsixum$ for purposes of visualisation.
Although there is a significant scatter, the median SEDs depicted as solid lines for both red and control QSOs show remarkably similar SEDs across the electromagnetic spectrum, starting to show differences only in the optical/UV regime consistent with the shape expected for red QSOs (Section \ref{sec:data}).

To understand in detail the origin of the SED diversity, in Figure \ref{fig:seds_Lz_descomp} we present average SEDs for each individual physical component, normalised at rest-frame 6 $\mu$m, and binned in groups of redshift and rest-frame 6 $\mu$m luminosities.
A quick visual inspection of the overall SEDs (grey solid and dashed lines) reveals the prevalence of flatter SEDs at lower redshift, where lower QSO luminosities are sampled, whereas SEDs at higher redshifts present `deeper' features in the near-infrared.
These deeper features are seen at high redshifts since the host galaxy contribution is no longer significant and does not fill the near-IR emission bands.
Indeed, a clear advantage of our SED decomposition approach is the ability to disentangle the host galaxy contamination which is more prominent at low-redshifts, and in particular in the low-luminosity bins.
Our study shows that for redshifts of $z<1.6$, the host galaxy emission can have a similar contribution to the near-infrared SED as the dusty torus and accretion disk components, in particular at luminosities of $\log \Lsixum \leq 45$.

A significant decrease of the relative contribution of the stellar population component with respect to the accretion disk emission is found as a function of redshift and luminosity.
In line with the average result in Figure \ref{fig:seds_median}, the median SEDs within different redshift and luminosity bins also show overall similar SEDs across the electromagnetic spectrum apart from the optical/UV regime.

A key result from the SED decomposition is the remarkable similarity between the torus SEDs for red and control QSOs, shown by the solid and dashed purple lines in Figure \ref{fig:seds_Lz_descomp}, across all redshift and luminosity bins.
This result hints towards no differences with respect to the dusty torus component between the control and red QSOs, suggesting the origin of the reddening does not arise from inclination.
A more detailed analysis of obscuration properties is carried out in Section \ref{subsec:obscuration}.

\subsection{Intrinsic accretion properties}\label{subsec:intrinsicprops}

\begin{figure*}
    \centering
    \includegraphics[trim={ 0.25cm 0.2cm 0.3cm 0cm},clip, width=0.8\linewidth]{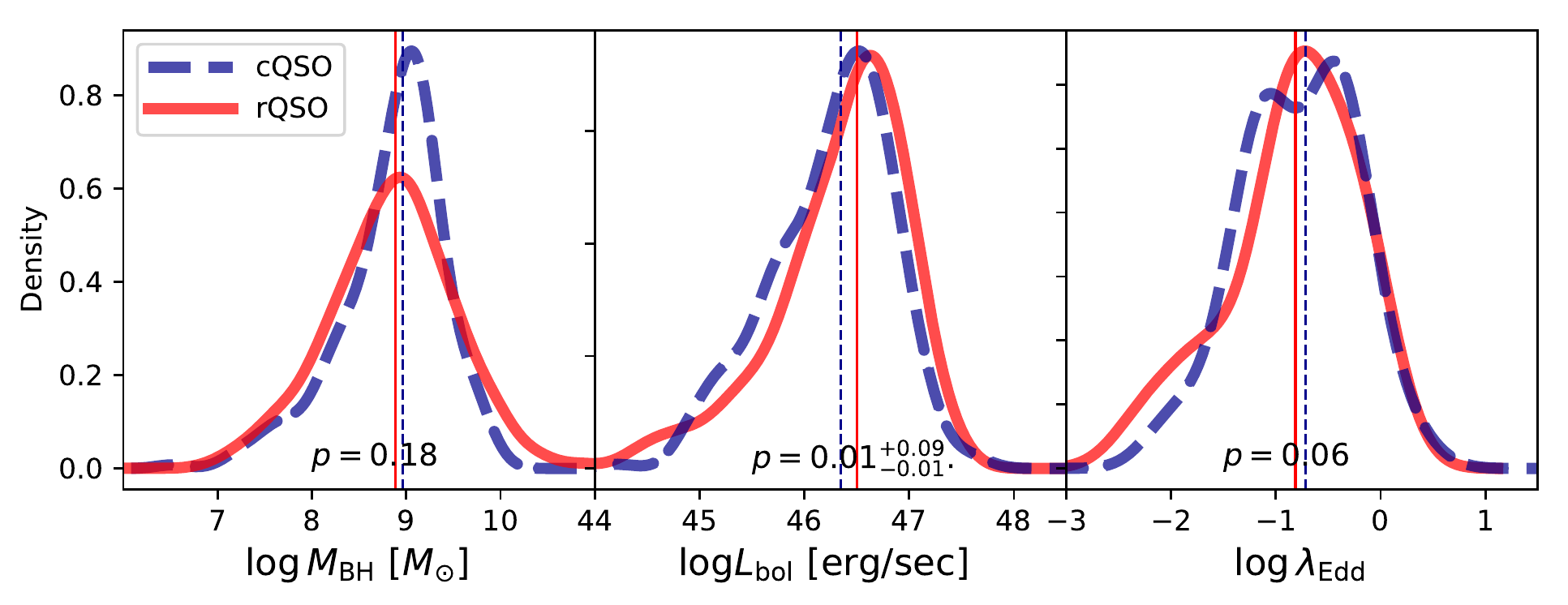}

    \caption{Distributions of black hole mass estimates ($\rm M_{\rm BH}$), reddening-corrected bolometric luminosities $\rm L_{\rm bol}$ and Eddington ratios $\lambda_{\rm Edd}$ for the control and red QSO samples. Only $\rm M_{\rm BH}$ estimates from spectra with quality tag 0 are considered, which comprise 90\% of our entire sample.  A Gaussian kernel-density estimator (KDE) with a bandwidth of 0.3 is used to display these distributions. With exception of the differences in the bolometric luminosity distributions, where significance is weak, no significant differences are found between the red and the control QSOs in these parameters. \gcr{The significance of the difference between the two distributions (KS-test $p$-value)  is reported below. The $p$-values were computed using the estimates reported by \citet{rakshit20} ($\rm M_{\rm BH}$ and $\lambda_{\rm Edd}$), and using the bootstrapped KS-test for the $\rm L_{\rm bol}$, where the posterior distributions from the MCMC-based SED fitting are compared and therefore uncertainties could be calculated. } }
    \label{fig:intrinsicprops}
\end{figure*}
To further test whether the optical-colour diversity used for the selection of our samples arise from intrinsic black hole accretion properties, rather than dust attenuation, we estimate black hole masses, bolometric luminosities, and Eddington ratios.
While estimates of these properties were provided for the entire SDSS catalogue by \citet{rakshit20}, these estimates have not been corrected for accretion disk reddening.
Our SED-fitting approach allow us to include these corrections and produce more reliable comparisons.

We estimate black hole masses for our sample adopting the same approach employed by \citet{rakshit20} and \citet{shen11}, with
\begin{equation}
   \log \left( \cfrac{M_{\rm BH}}{M_{\odot}} \right)= A+B \log \left( \cfrac{ \lambda L_{ \lambda}}{10^{44}\, \rm erg\,  s^{-1}} \right) + 2 \log \cfrac{FWHM}{\rm km\,s^{-1}},
\end{equation}
where $ \lambda L_{\lambda}$ are the monochromatic luminosities L$_{1350}$, L$_{3000}$, L$_{5100}$, estimated from the reddening and host galaxy corrected SEDs.
The FWHMs are obtained from [C~{\sc iv}], [Mg~{\sc ii}] and H$\beta$ lines, reported by \citet{rakshit20} for the SDSS QSOs at $z<0.8$, $0.8<z<1.9$, and $1.9<z<2.5$ respectively. 
The constants A and B\footnote{A,B = (0.910,0.50) for H$\beta$; A,B = (0.860,0.50) for [Mg~{\sc ii}]; A,B=(0.660,0.53) for [C IV]} are given by \citet{rakshit20} as well, where they use corrections based on virial estimates to the corresponding lines inferred by \citet{vestergaard06} and \citet{vestergaard09}.
We note that the monochromatic luminosities from our SEDs agree with those reported by \citet{rakshit20} for the control and red QSOs before the reddening correction is applied. 
After the reddening correction, the monochromatic luminosities of red QSOs are significantly larger than those reported by \citet{rakshit20}, as expected.
While red QSOs would have lower median M$_{\rm BH}$ values than the control sample based on black hole masses reported by \citet{rakshit20} ($\Delta \log \rm M_{\rm BH} \sim 0.25$), this effect disappears after applying the reddening correction  ($\Delta \log \rm M_{\rm BH} < 0.1$). 
In Figure \ref{fig:intrinsicprops} both populations have similar distributions with almost identical median M$_{\rm BH}$ values at $\rm \log M_{\rm BH} \sim 9$\,\Msun\, and a slight difference in the width of the distributions.
These results are in line with the emission line width and luminosity analysis presented by \citep{klindt19}, where they investigate the intrinsic accretion properties by comparing line width distributions for blue, control and red QSO, finding no significant differences.

\gcr{Bolometric luminosities are also estimated from the integrated emission from the accretion disk component (BBB)  over the wavelength range of 500 $\AA$ to 1\,\micron, with a small additive correction factor of $\delta \log \rm L_{\rm bol} = 0.3$ to account for X-ray emission not included in the SEDs.
The bolometric luminosities reported by \citet{rakshit20}, on the other hand, are computed using a bolometric correction to the monochromatic luminosity at 3000 $\AA$, following $\rm L_{\rm bol} = 5.15 \times L_{3000}$ \citep{shen11}.
We find that our SED-based bolometric luminosity distribution is equivalent to that reported by \citet{rakshit20} for the entire sample before the reddening correction is applied.
After the application of the reddening correction, however, our estimates for the red QSOs are significantly larger, as expected.}
The second panel of Figure \ref{fig:intrinsicprops} shows the $\rm L_{\rm bol}$ distributions, where slightly larger bolometric luminosities for red QSOs can be recognised. 
A bootstrapped KS-test (Section \ref{subsec:stats}) shows these differences are small though statistically significant (p-value = $0.01^{+0.09}_{-0.01}$).
We explore these implications in Section \ref{subsec:obscuration}.

Based on our estimates of $M_{\rm BH}$ and $L_{\rm bol}$ we calculate Eddington luminosities and ratios for our comparative samples of red and control QSOs in the right panel of Figure \ref{fig:intrinsicprops} following,

\begin{equation}
    L_{\rm Edd}= 3.28 \times 10^{4} \left( \cfrac{M_{\rm BH}}{M_{\odot}} \right)
    L_{\odot} \\
    \lambda_{\rm Edd} = \cfrac{L_{\rm bol}}{L_{\rm Edd}}.
\end{equation}

Once we have accounted for host galaxy and dust reddening corrections, we find strikingly similar distributions of black hole masses and Eddington ratios for red and control QSOs.

\gcr{Both QSO populations have median Eddington ratios around $\lambda_{\rm Edd}= 0.1$, suggesting that the intrinsic accretion properties of the SDSS QSOs included in this study are very similar for both red and control QSOs}.
Based on these observations, we conclude that we find no differences in the intrinsic accretion properties of red and control QSOs.

\subsection{Obscuration properties of blue and red QSO}\label{subsec:obscuration}

\begin{figure}
    \centering
    \includegraphics[trim={ 0.2cm 0.2cm 0.2cm 0cm},clip, width=\linewidth]{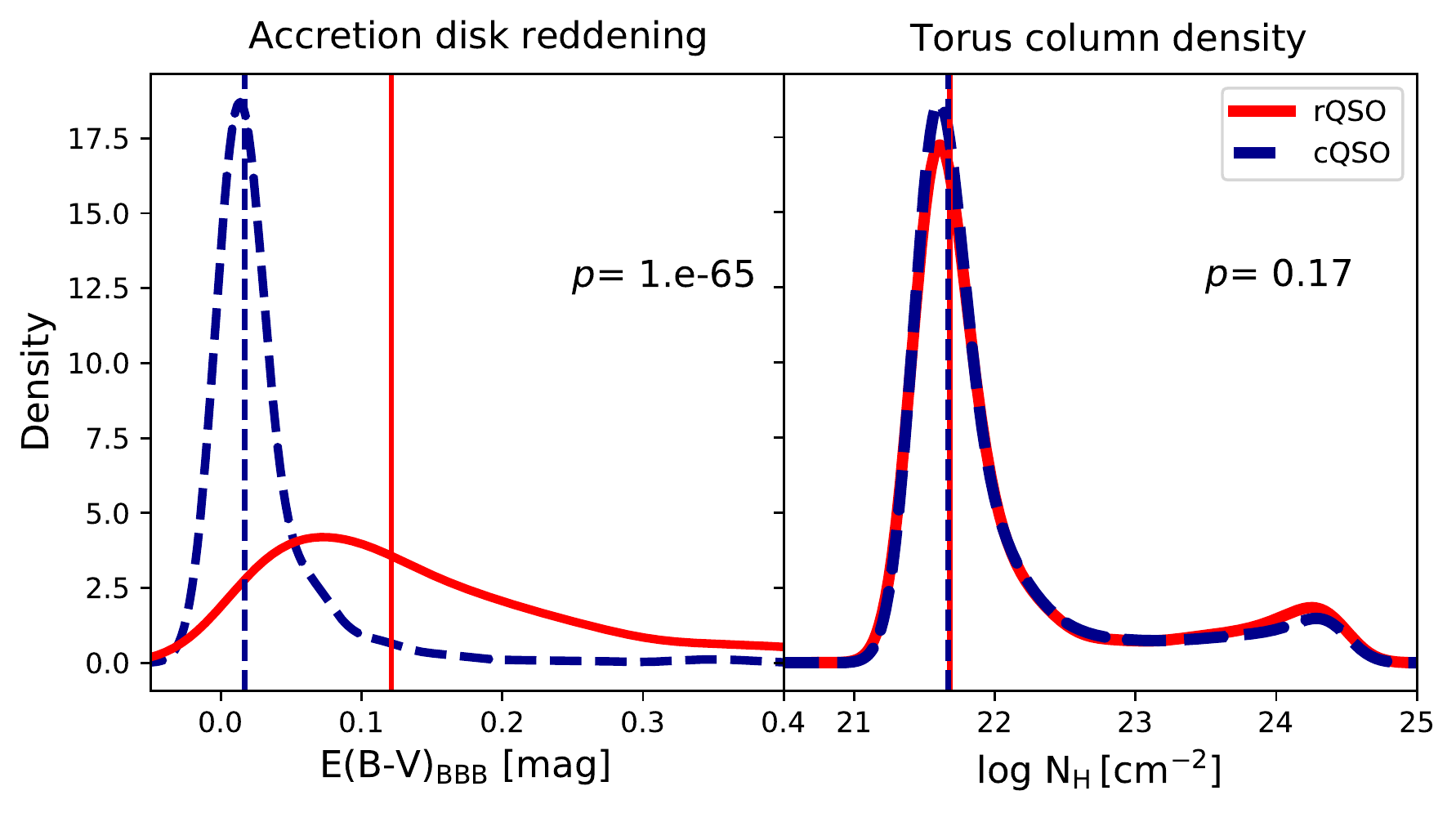}
    \caption{Probability distributions of the accretion disk reddening parameter E(B-V), and the column density $\rm N_{\rm H}$. The E(B-V) parameter was inferred assuming the SMC attenuation law by \citet{prevot84} on the accretion disk emission, while the $\rm N_{\rm H}$ is te quantity that parametrizes the (X-ray-normalised) infrared torus templates by \citet{silva04}.  These distributions are constructed as the superposition of 100 random draws from each source. A gaussian kernel-density estimator (KDE) with a bandwidth of 0.1 is used to display these distributions. }
    \label{fig:distributions_ebv_nh_r}
\end{figure}

To investigate the origin of the red colours in our QSO sample we focus on two physical components of our SED fitting model: the reddened accretion disk emission itself, and the reprocessed emission by hot dust at nuclear scales commonly modelled with a torus structure.
Two output quantities that parametrise these components are the dust extinction parameter that attenuates the blue part of the accretion disk emission, E(B-V)$_{\rm BBB}$, and the column density of the obscuring medium inferred from the torus infrared SED, $\rm N_H$.
In Figure \ref{fig:distributions_ebv_nh_r} we show the posterior distributions of the dust attenuation of the accretion disk SED E(B-V)$_{\rm BBB}$ for red QSOs and control QSOs.
The posterior distributions are reconstructed by the superposition of 100 draws from the PDF of each source.

As expected by their selection, we find a significant difference in the distributions of E(B-V)$_{\rm BBB}$ for red and control QSOs (bootstrapped KS-test p-value$\sim 10^{-65}$), where red QSOs have E(B-V)$_{\rm BBB} = 0.12^{+0.21}_{-0.08}$ mag (A$_{\rm V} = 0.32^{+0.57}_{-0.21}$ mag), and control QSOs have E(B-V)$_{\rm BBB} = 0.02^{+0.04}_{-0.01}$ mag (A$_{\rm V} = 0.05^{+0.11}_{-0.03}$ mag).

In contrast to what is expected from the dust attenuation, we find that both red and control QSOs have strikingly similar distributions of dust column density $\rm N_H$ as measured by the torus emission component (Figure \ref{fig:distributions_ebv_nh_r}). 
The bulk of both populations ($\sim 83\%$) show low column densities of $\rm \log N_{\rm H} < 22.$, with no significant differences (bootstrapped KS-test $p$-value= 0.17).
This is in line with the similar shapes of the dusty torus SED components found among red and control QSOs shown in Figure \ref{fig:seds_Lz_descomp}, suggesting that the main infrared emitting component is similar for the two populations.
We remark that the  $\rm \log N_{\rm H}$ values quoted here are not inferred from the X-rays but from the infrared alone, using the $\rm \log N_{\rm H}$ term to parametrize the torus templates by \citet{silva04}.
These estimates of N$_{\rm H}$ are, however, overall consistent with measures from X-rays spectroscopic studies of SDSS QSOs \citep{wilkes02, urrutia05}, although these are found to be significantly higher for red QSOs selected in the near-infrared \citep{goulding18, lansbury20}.

The similarity in the shape of the torus SED suggests that the nuclear dust structure, which is responsible for the AGN unification scheme, has no direct connection with the dust attenuation that characterises the red QSO population.
Indeed, would the reddening arise from a viewing angle effect, in particular from viewing angles closer to an edge-on view of the dust structure (Type 2 QSO), this would imprint the torus SED shape due to the higher observed column density. 
The lack of diversity of the torus SEDs for all red and control QSOs, as inferred by our model, suggests that the torus viewing angle effect is not causing the reddening.
One systematic limitation that would synthetically produce this observation is the possibility that the torus emission templates used for the SED fitting do not recover the effect of torus inclination properly.
However, our previous work has shown that this is not the case.
Using the same machinery and semi-empirical templates \citep{silva04} on spectroscopically confirmed Type 1 and Type 2 QSO, in \citet{CR16} we demonstrated that these models successfully recover a diversity of torus structures.
In particular, these were able to recover the spectroscopic Type 1 and Type 2 QSO classification based solely on the shape of the SEDs, using the diversity found in the $\rm N_{\rm H}$ and E(B-V) parameters as the drivers of the classification.
Additionally, the column density distributions and torus SED shapes observed in Type 1 QSOs in the COSMOS field by \citet{CR16}, are fully consistent with the distributions observed by the SDSS QSOs in this work, independently of colour.

In the following sections we explore two scenarios that explain the lack of differences in the infrared emission of red and control QSO, despite the clear difference in the dust-attenuated optical and UV emission.
Firstly, we investigate the dust reprocessing efficiency in Section \ref{subsubsec:covfactor}.
Secondly, in Section \ref{subsubsec:residuals} we investigate whether there is emission that is not recovered by our models by computing the infrared residuals from the SED fitting.

\subsubsection{Reprocessing efficiency}\label{subsubsec:covfactor}
\begin{figure*}
    \centering
    \includegraphics[trim={ 0.15cm 0.2cm 0.2cm 0cm},clip, width=0.49\linewidth]{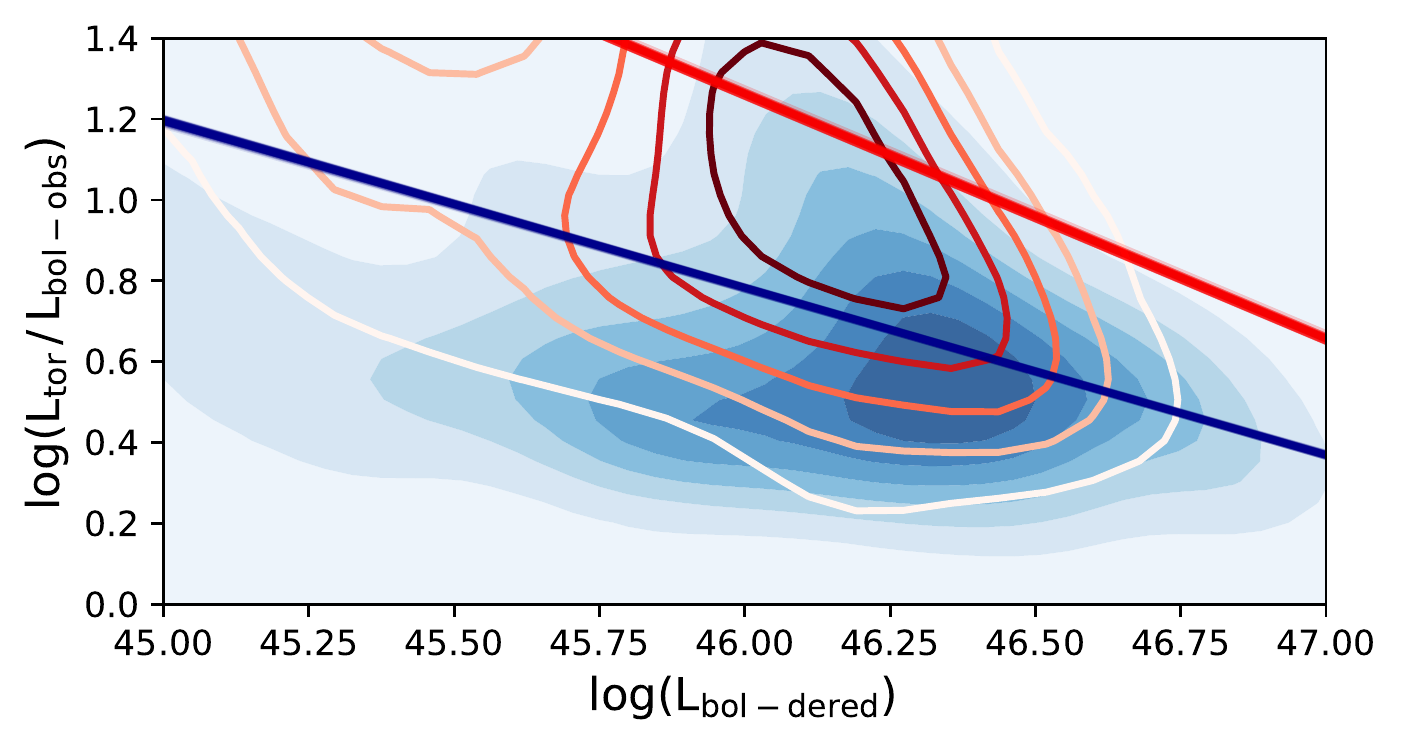}
    \includegraphics[trim={ 0.15cm 0.2cm 0.2cm 0cm},clip, width=0.49\linewidth]{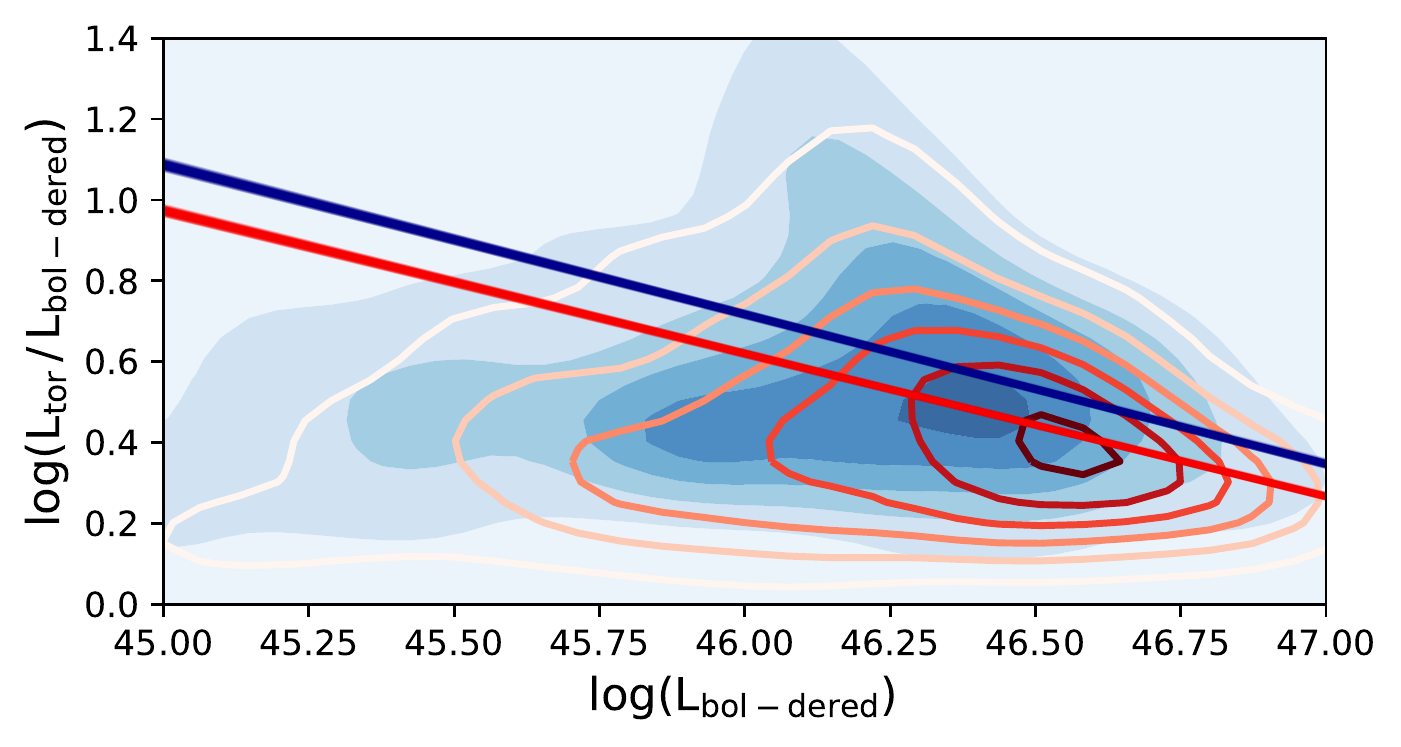}
    \includegraphics[trim={ 0.15cm 0.2cm 0.2cm 0cm},clip, width=0.49\linewidth]{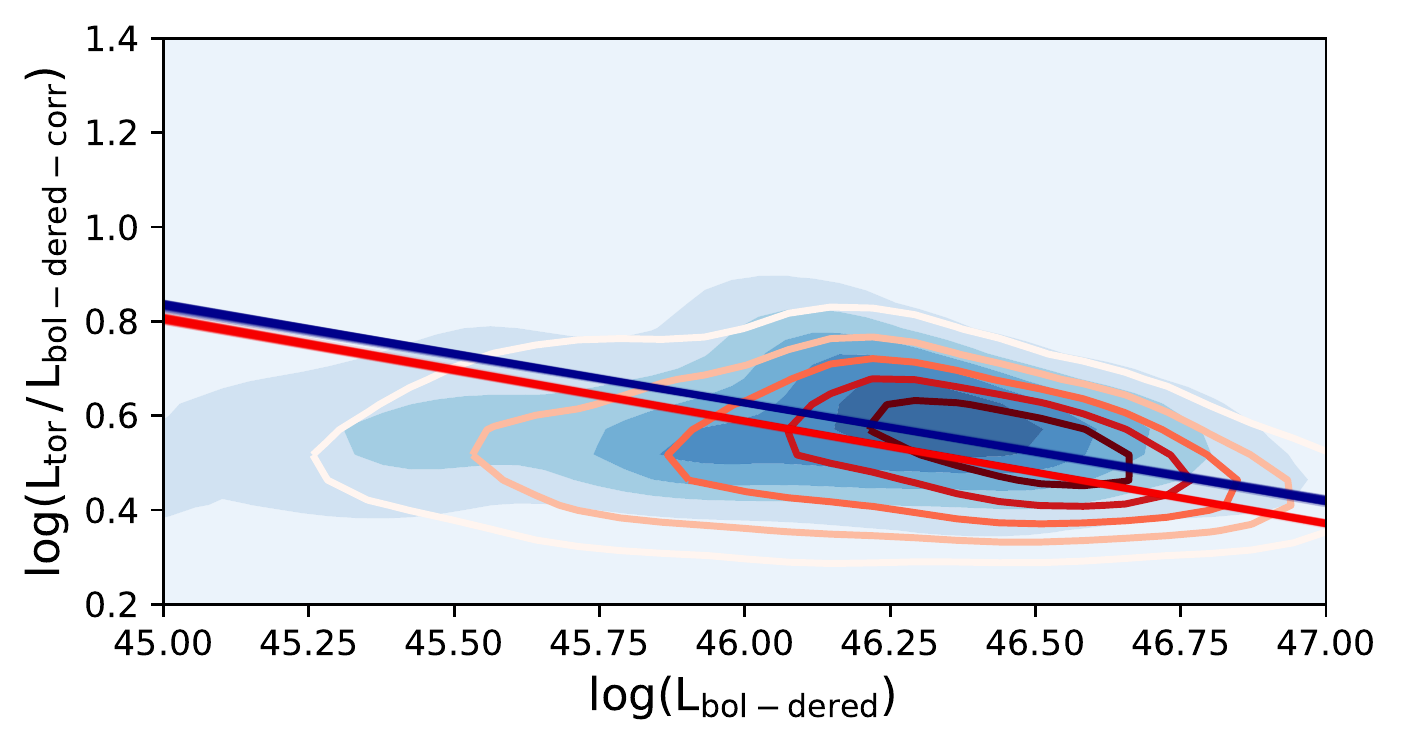}

    \caption{ We show composite two-dimensional posterior distributions of the torus reprocessing efficiency, defined as the ratio of torus luminosities and  bolometric luminosities, plotted as a function of bolometric luminosities. These are constructed by sampling 100 realisations from the posterior of each source to properly account for the uncertainties. The red contour lines (for red QSOs) and the blue contour areas (for control QSOs) range between 20 and 100 per cent, in intervals of 20 per cent. In the upper panels we show this ratio for the observed bolometric luminosities (left panel) and for the reddening-corrected (intrinsic) bolometric luminosities (right panel)  for red and control QSOs.  The lower panel shows the torus reprocessing efficiency corrected for the torus anisotropy following \citet{stalevski16}, to provide a better description of the covering factor.}
    \label{fig:RvsLbol}
\end{figure*}
In Figure \ref{fig:RvsLbol} we investigate the fraction of the intrinsic bolometric AGN luminosities that is reprocessed by the dusty torus and re-emitted in the infrared, i.e the torus reprocessing efficiency, which is used as a proxy for the covering factor of the obscuring medium \citep{maiolino07,treister08, lusso13, stalevski16}.
We compute the AGN bolometric luminosities $\rm L_{\rm bbb}$ and torus luminosities $\rm L_{\rm tor}$  by integrating over the fitted templates of the accretion disk and torus physical components, covering the range between 0.05 and 1 \micron, and 1 \micron\, and 1000 \micron, respectively. 
We note that our results do not change significantly if the range 0.1 and 1 \micron\, is used for the accretion disk emission.
We also stress that the SED decomposition method we use to compute the $\rm L_{\rm bbb}$ is more robust than bolometric luminosities scaled up from one or a few wavelength measurements using a single bolometric correction, since more information/data is included in the estimation. 

We compute the torus reprocessing efficiency as the ratio of the accretion disk emission (bolometric luminosity) and torus luminosities, considering both the observed accretion disk luminosity ($\rm L_{\rm tor}/\rm L_{\rm bbb-obs}$ Figure \ref{fig:RvsLbol}, left) and the reddening-corrected one ($ \rm L_{\rm tor}/\rm L_{\rm bbb-dered}$, Figure \ref{fig:RvsLbol}, right).
This ratio is related to the covering factor, which is in turn connected to the probability to observe an AGN as obscured or unobscured \citep{lusso13, stalevski16}.
We remind the reader that, in contrast to other SED fitting models, \textsc{AGNfitter} models the accretion disk and torus components independently, allowing to investigate the variations in this ratio.

\gcr{In the left upper panel of Figure \ref{fig:RvsLbol}, we show this ratio as a function of the intrinsic accretion disk luminosities (reddening corrected) for our QSO samples.}
A clear difference between red and control QSOs can be recognised, which is expected from the dust attenuation in red QSOs. 
We next test whether this observation holds after correcting the bolometric luminosities for reddening of the accretion disk.
In the right panel of Figure \ref{fig:RvsLbol}, we show this ratio as a function of the intrinsic reddening-corrected accretion disk luminosities. 
We find that the distribution of the $\log \rm L_{\rm tor}/L_{\rm bol}$ has significantly changed after the reddening correction, now around physical values below unity, and that red QSOs are more similarly distributed to the control QSOs.
We find that the median value of the torus reprocessing efficiency is $0.437 \pm 0.002$ for red QSOs, which is lower than that observed for the control sample, at values of $0.550 \pm 0.002$, where uncertainties are bootstrap errors.
The linear fits reveal slope values of -0.35 for red QSOs and\,-0.37 for the control sample.
These values reflect an apparent negative evolution of the obscuration in both QSO populations as a function of bolometric luminosity, consistent with the receding torus scenario, where the increasing radiation pressure of higher QSO luminosities may push the obscuring medium away from the luminous accretion disk (e.g. \citet{simpson05}, but also note caveats and different interpretations in \citet{netzer16, stalevski16}).
We perform a non-parametric two-sample KS statistic  described in Section \ref{subsec:stats} to test the hypothesis that the two distributions of red and control QSOs are drawn from the same distribution of reprocessing efficiency. 
We find we can reject this hypothesis with a confidence level of 99.9 per cent (p-value= 2$\times {10^{-5}}$) and therefore conclude this difference is statistically significant.
The tentative observation, supported by the KS-test, that red QSOs have lower reprocessing efficiency would hint towards a slight difference in the torus composition, possibly related to larger gas to dust fractions and a more diffuse obscuring medium \citep{lansbury20}.

Finally, we estimate the torus covering factor by accounting for the torus anisotropy and its effect on  our estimates of the reprocessing efficiency.
Based on a 3D radiative transfer model, \citet{stalevski16} quantified this effect, showing that its influence can be non-linear and strongly dependent on the assumed torus optical depth. 
Motivated by the output on torus column densities from the SEDs, we assume a torus optical depth of $\tau=3$ for both red and control QSO samples, and apply the corrections reported by \citet{stalevski16} to account for the torus anisotropy.
We show the resulting relation in the lower panel of Figure \ref{fig:RvsLbol}. 
The differences between distributions of red and control QSOs decrease after this correction, however they remain significant (p-value = $2 \times 10^{-5}$ ).
Nonetheless, whether it is indeed the case that red QSOs have less efficient torus obscuration, or if we adopt the more conservative interpretation that both populations have equivalent efficiencies based on the weak differences,  none of these scenarios can explain the origin of the red QSO colours, reinforcing our previous hypothesis that no connection exists between QSO reddening and their dusty tori.

Possible systematic effects, such as a uncertainties in choosing appropriate extinction curves to describe QSO reddening might affect our results, especially where no data points are available to constrain the curve.
However, the E(B-V) distributions from the \citet{prevot84} model are consistent with what has been found in other studies, and the quality of the fits show SMC laws are a good description of the reddening in these sources \citep[e.g.][ and Fawcett et al. in prep.]{hopkins04, glikman12}.
Finally, as mentioned above, we verified that a choice of a different wavelength range (0.1--1 $\mu$m, instead of 0.05--1$ \mu$m) does not change our results significantly.
Another possible systematic effect is the choice of intrinsic accretion disk emission model, which could over-predict the intrinsic luminosities for red QSOs.
We expect that the effect of such a scenario is not significant, since the intrinsic accretion model of \citet{richards06} performs well for the control sample, and no differences in the intrinsic accretion properties are expected based on our spectral analysis in Section \ref{subsec:intrinsicprops} (Figure \ref{fig:intrinsicprops}).
Overall, we find the infrared reprocessing efficiency of red QSOs is slightly lower or at least comparable to the control QSO sample, suggesting the torus structure is disconnected to the reddening properties of red QSOs.

\begin{figure*}
    \centering
    \includegraphics[trim={ 0.2cm 0.2cm 0.2cm 0cm},clip, width=\linewidth]{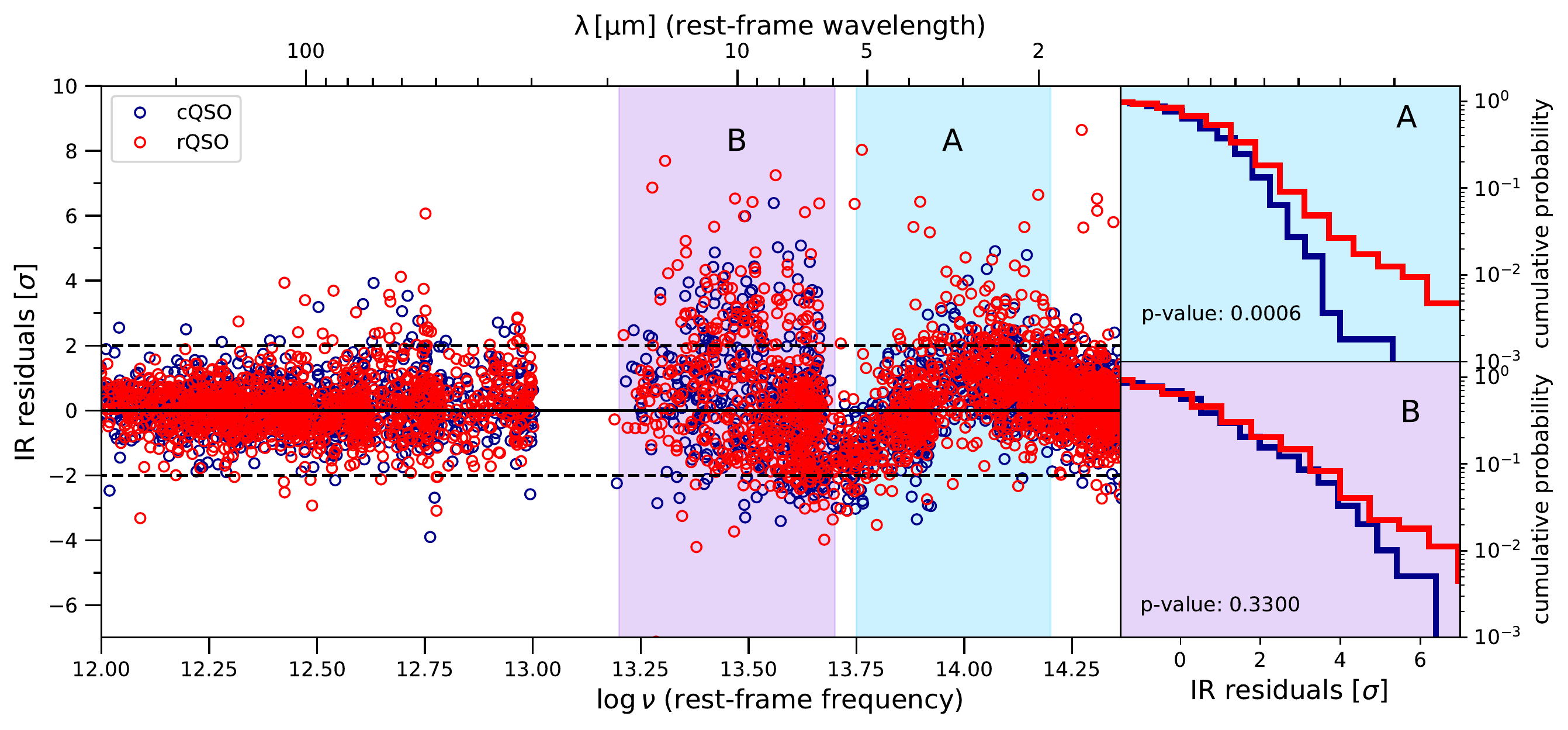}

    \caption{Residuals from the best fits for each source as a function of rest-frame frequency. Two small `bumps' are identified in the MIR regime, where the torus models included in our model show the largest discrepancy with the data. The main discrepancies have positive residuals meaning the model cannot reproduce a part of the flux at these frequencies. In the right panels we show cumulative distributions of the incidence of residuals in each sample. We investigate whether these residuals are connected to the optical colour of the QSO, despite the fact these bands are independent in the fitting. We find that red QSOs have larger MIR emission excess compared to the model in the regime of $13.75<\log \, \nu<14.2$ (2-5 $\mu$m), showing significant residuals $(> 2 \sigma )$ in 20 per cent of the cases, in contrast to the 10 per cent in the case of control QSOs.} 
    \label{fig:residuals}
\end{figure*}

\subsubsection{Residual dust distributions}\label{subsubsec:residuals}
\begin{figure}
    \centering
    \includegraphics[trim={ 0.2cm 0.2cm 0.3cm 0cm},clip, width=0.9
    \linewidth]{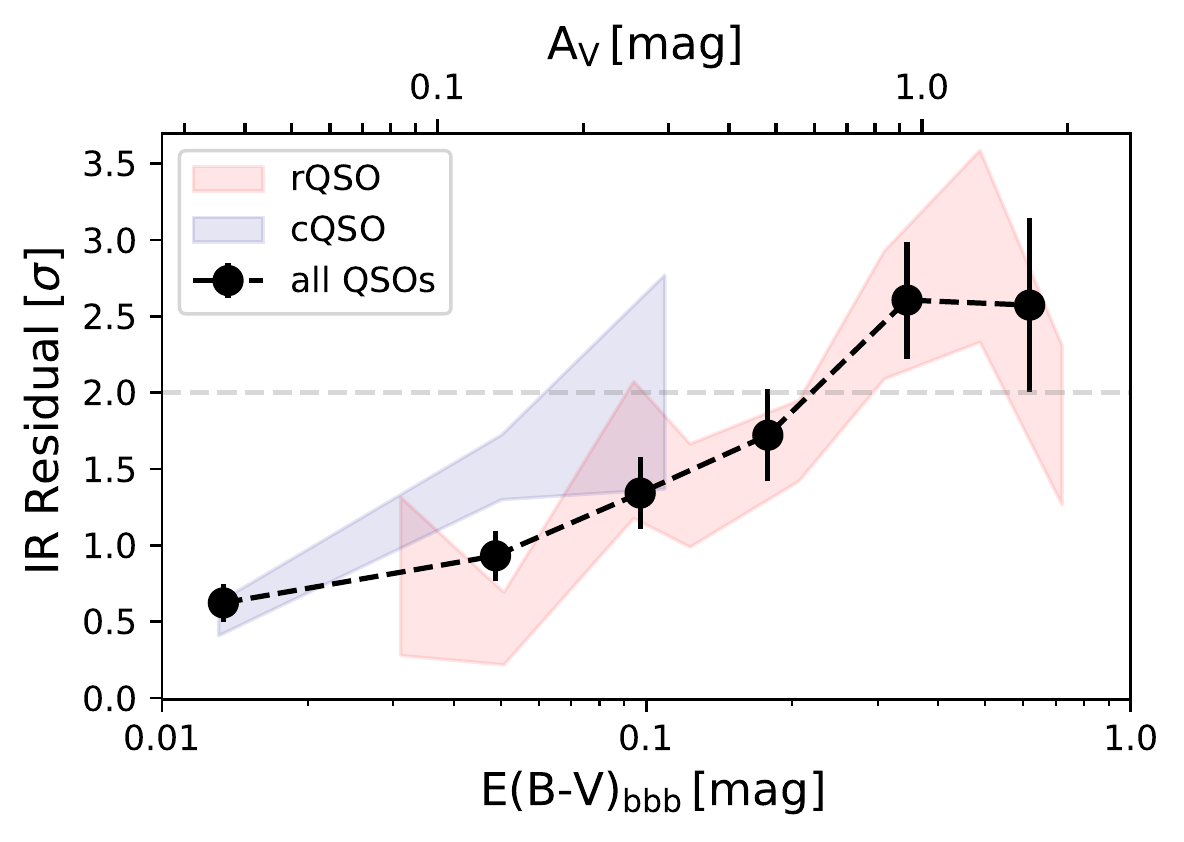}
    \caption{Median infrared residuals are shown as a function of the QSO reddening parameter, E(B-V)$_{\rm BBB}$. Black circles represent the median residual values for each E(B-V)$_{\rm BBB}$ bin for all QSOs irrespective of colour  classification. The error bars represent bootstrap 1-$\sigma$ errors.  The same relation is shown for the red and control QSOs as red and blue shaded areas, respectively. A clear trend can be observed, where the infrared residuals increase with larger reddening values for all samples. }
    \label{fig:res_ebv}
\end{figure}

We now investigate the existence of infrared features that could be linked to the dust attenuation in red QSOs but are not recovered by torus  models.
To quantify such emission we compute the residuals of the best fits for the QSO SEDs in the infrared, expressed in terms of significance given the data uncertainties.
We plot these as a function of rest-frame $\log$ frequency in Figure \ref{fig:residuals}.
Although the residuals are overall small and not significant in the majority of the frequencies covered, some structure can be recognised in the mid-infrared regime. 
As expected, the SEDs do not have coverage in the rest-frame wavelength regime between $\lambda \sim 20-30$ \micron.
Interestingly, two peaks are recognised in the regions highlighted with colours in Figure \ref{fig:residuals} where significant positive residuals are found ($\sigma$>3).
Positive residuals mean that there is emission from the data which cannot be recovered by the models.

We investigate the origin of these peaks and compute the incidence of residual luminosities for each QSO sample in the right side panels of Figure \ref{fig:residuals}.
We find that there is a larger incidence of residuals in the SEDs of red QSOs as compared to control QSOs, which is particularly present in the infrared regime of rest-frame $\log \nu \sim $14 Hz (2-5 $\mu$m; panel A).
A KS test on this difference reports a p-value of 0.0006, suggesting the enhanced flux excess in red QSOs is highly significant. 
We further evaluate the fraction of sources with residuals above this value.
We find that 20 per cent (54/306) of the red QSO sample have residuals of $\sigma$>2, whereas the fraction reduces to 11 per cent (35/306) for the control QSO population. 
A few examples of red QSO SEDs which clearly present these residuals were also included in Figure \ref{fig:SEDs} (two lower left SEDs).

A slight enhancement of infrared residuals can be recognised as well in the rest-frame $\log \nu \sim 13.5$ Hz regime ($\lambda \sim 10$ \micron), albeit with lower significance.
With black body temperatures at $\sim 150-500$ K, the residuals in this longer-wavelength regime suggest that warm dust emission contributions might be present for Type 1 QSO, irrespective of colour. 
For the purposes of this investigation we restrict our further analysis to the infrared residuals in the 2-5 $\mu$m regime.

\begin{figure*}
    \centering
    \includegraphics[trim={ 0.2cm 0.2cm 0.3cm 0cm},clip, width=0.8\linewidth]{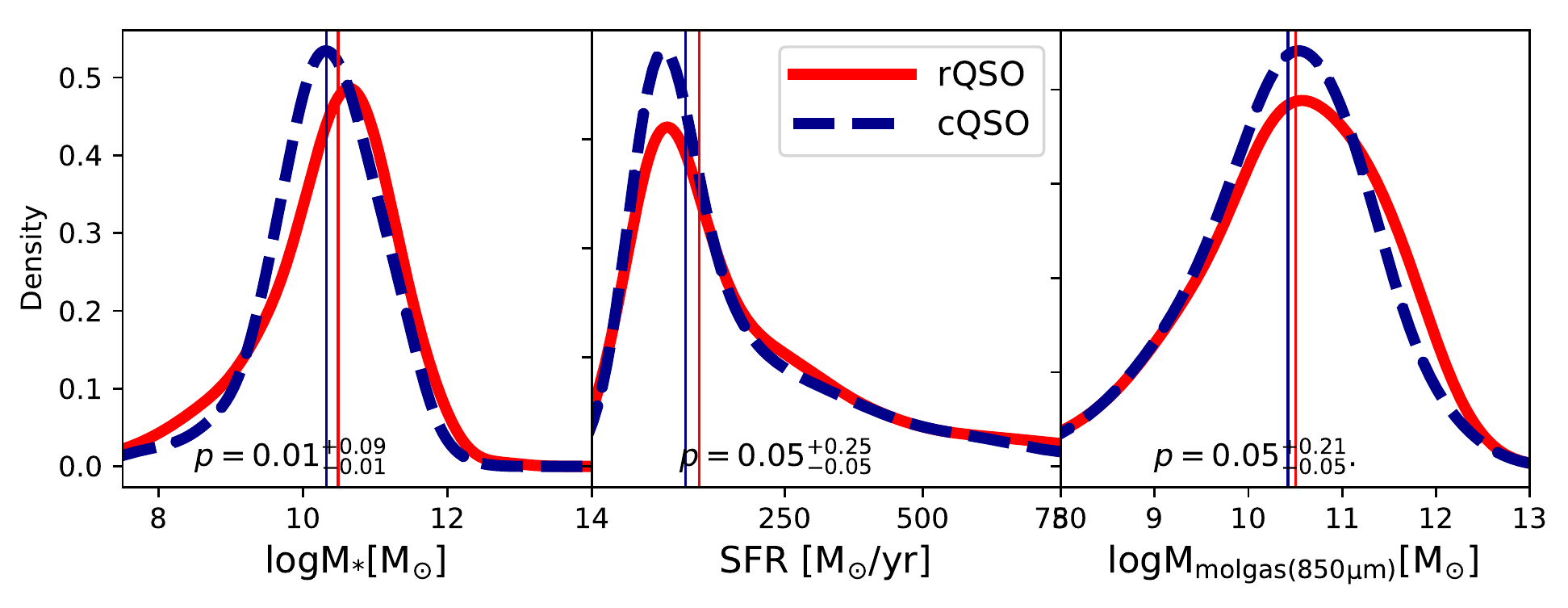}
    \caption{Composite posterior distributions of host galaxy parameters inferred through SED-fitting. These are constructed by sampling 100 realisations from the posterior distributions of each source, and we apply a KDE with a bandwidth for 0.3 for display as explained in detail in Section \ref{sec:data}.  Stellar masses are estimated based on the modelling of near-infrared and optical emission with stellar population models \citep{bruzual03}. Star formation rates are computed based on  total infrared luminosities (L$_{IR(8-1000\mu \rm m}$) from the modelling of the infrared photometry with cold dust models \citep{schreiber18}. Molecular gas masses are estimated from monochromatic luminosities of the dust continuum SED at $850\mu \rm m$ and the empirical relation by \citet{scoville16}. No significant differences are found between the distributions of red and control QSOs for these host-galaxy parameters.}
    \label{fig:distributions_hostgalaxy}
\end{figure*}
We further investigate a direct relation between the incidence of infrared residuals in the 2-5 $\mu$m regime as a function of QSO reddening in Figure \ref{fig:res_ebv}.
A clear trend can be observed, where the infrared residuals increase with larger reddening values for all samples, in spite of the fact that both parameters ( E(B-V)$_{\rm BBB}$ and residuals) are determined by independent regions of the SED (rest-frame near-infrared  and optical/UV, respectively).
This observation offers additional support to a potential link between this excess hot dust emission and the colours in red QSOs.
It is important to note that the fact that the infrared residuals are characterised by low levels of $\sigma$ is mainly determined by the low signal-to-noise of the infrared photometric measurements which is our only means to parametrize such excess. 

Excess emission at this wavelength regime ($2-5\,\mu$m) corresponds to dust temperatures of 550-1700K, assuming a black body and applying Wien's law.
In the near-infrared regime, we should also consider the potential contribution from the host galaxy.
However, we can rule out this possibility for the following reasons. 
First, based on the general SED of galaxies it is highly unlikely that the host galaxy SED presents a peak around $2-5\,\mu$m, since the reddest possible peaks for the most massive, older and cooler stellar populations  are located at wavelengths bluer than 1.5 $\mu \rm m$.
Instead, galaxy SEDs show overall flat SEDs \citep{bruzual03} in the wavelength regime relevant here, being unable to  
explain the red excess bumps around $\lambda \sim 4\, $\micron.
Most crucially, the overall galaxy contribution is negligible in the majority of our sources due to their high-luminosities, as seen in Figure \ref{fig:seds_Lz_descomp}.

We conduct a few exercises to test the robustness of this finding.
First we investigate whether the higher infrared excess in red QSOs is driven by a fraction of sources at a specific redshift.
We find this is not the case, finding a clear enhancement of infrared residuals for red QSOs compared to the control samples across most redshift ranges covered by the study, with p-values (for the respective redshift ranges) of 0.01 ($0.2<z<0.9$), 0.16 ($0.9< z<1.6$), 0.02 ($1.6< z<2.2$), and  0.001( $2.2\leq z<2.5$).
We further test whether this excess is driven by the observations of a specific instrument (WISE or IRAC).
Although we find a significant enhancement in both samples which include WISE+IRAC (northern fields) and those with only WISE data available (GAMA fields), it is clear that those with deeper measurements and a better sampling of the SED around  rest-frame 3 $\mu$m have excess detections of higher significance.
In particular, since the bulk of the QSO population is located at $z>1$ the coverage of the IRAC channels ch3 (5.8 $\mu$m) and ch4  (8.0 $\mu$m) is crucial for detecting this excess. 
We further test whether this emission could be recovered by more complex state-of-the-art torus models which include other free parameters such as opening angle of the torus, and different dust grain sizes.
In particular, we perform a SED-fitting test run on a small subsample of our data with significant residuals using the torus model by \citet{stalevski16}, finding no improvement in the fit of those features.

Given that these infrared residuals cannot be reproduced by torus SED models we investigate whether these have originated from additional hot dust emission components, such as the presence of dusty winds at radii close to the sublimation radius.
We discuss this scenario in Section \ref{sec:discussion}.
In particular, hot dust models exist which  include the emission from polar cones of dusty winds lifted up from the inner parts of the torus by radiation pressure \citep{honig17, costa18, venanzi20}. 
We note that including these models in the SED fitting implies sampling highly complex parameter spaces \citep[11 parameters, ][]{honig17}.
This is out of the scope of this investigation given the limitations of the available infrared photometry and the lack of MIR spectroscopic data.

\subsection{Host galaxy properties of red and control QSOs}\label{subsec:galaxies}

Finally, we also investigate a possible link between the properties of the host galaxies and the reddening of QSOs.
In particular, we focus on the star-formation rates, stellar masses, and molecular gas masses estimated from the SED decomposition technique.
We do not include parameters such as the galaxy age or galaxy dust attenuation in our analysis since these are constrained based on the bluer parts of the SED which is largely dominated by the QSO emission, making these estimates  highly unreliable.
Following the same argument, we note that the SFR estimates presented in this work are \textit{obscured} star-formation rates, i.e. these have been estimated solely based on the integrated luminosities of the infrared regime between $8-1000$ \micron, by fitting the photometric data with the \citet{schreiber18} cold dust model.
Although estimating stellar masses from photometry for Type 1 QSOs can be a challenging endeavour, the near-infared data has significant constraining power especially for moderate QSO contributions and in the lower redshifts bins.
Based on our knowledge of the QSO accretion disk emission constrained by the blue optical bands and UV, and supporting the inference with prior information on expected host galaxy luminosities (see Section \ref{subsec:agnfitter}), we are able to disentangle the host galaxy component and derive estimates of the stellar masses and their associated uncertainties.
We note, however, that in the higher QSO luminosity and high redshift bins ($\log \Lsixum$>44.6\,\ergs, $z>1.$), where the host galaxy emission starts becoming less important compared to the QSO luminosities, the uncertainties are expected to increase, and we rely on our Bayesian technique to recover these as demonstrated by the statistical tests presented in \citet{CR16}. 
Molecular masses are calculated using the empirical calibration of the Rayleigh-Jeans luminosity-to-mass ratio by \citet{scoville16}, based on the SED-estimated rest-frame 850 \micron \, luminosities and associated uncertainties.

In Figure \ref{fig:distributions_hostgalaxy} we examine the composite posterior probability density functions of  the relevant parameters:  stellar masses, infrared star-formation rates, and molecular gas masses.
We construct these composite PDFs, again by combining 100 draws from each source's PDF and apply a Gaussian KDE for display.
The overall stellar mass distribution for red QSOs has a median value of $\log \rm M_{*}(rQSO) = 10.48^{+0.69}_{-1.04}$ \Msun\, (where uncertainties correspond to the 14th and 86th percentiles) whereas control QSOs have $\log \rm M_{*}(cQSO) = 10.32^{+0.72}_{-0.72}$ \Msun.
To assess the significance of the difference between these two distributions we explore the probability that these have been drawn from the same distributions, using the KS-test described in Section \ref{subsec:stats}.
We find a median p-value = $0.01^{+0.09}_{-0.01}$, suggesting the differences in stellar masses are significant, although marginally given the uncertainties in the p-value.
In order to further test the SED-based stellar masses, we compare these to those estimated based on the black-hole-mass-stellar-mass relation, empirically calibrated by \citet{bennert11} as a function of redshift.
The M$_{\rm BH}$-based stellar mass distribution for the QSO samples have median values of $\log \rm M_{*}(rQSO) = 10.82^{+0.52}_{-0.57}$ \Msun \, and $\log \rm M_{*}(cQSO) = 10.89^{+0.32}_{-0.47}$\,\Msun.
This shows that, although the SED-inferred stellar masses are overall consistent with the M$_{\rm BH}$-based estimates within the uncertainties, the M$_{\rm BH}$-based estimates do not recover the slightly enhanced stellar masses for red QSOs reported above. 
We therefore conclude that the overall stellar mass distributions show no difference between red and control QSOs.

The SFR distributions are also remarkably similar, where the QSO samples have median values of $\rm SFR_{\rm IR}(rQSO) = 95^{+364}_{-85}$ \Msun yr$^{-1}$ and $\rm SFR_{\rm IR}(cQSO) = 70^{+309}_{-62}$ \Msun yr$^{-1}$.
Here we find that the distributions are not significantly different with a median p-value = $0.05^{+0.25}_{-0.05}$.
Similarly, the molecular masses have median values of  $\log \rm M_{\rm mol}(rQSO) = 10.5^{+0.9}_{-1.1}$ \Msun\, and  $\log \rm M_{\rm mol}(cQSO) = 10.4^{+0.8}_{-1.0}$ \Msun\,.
We find that the distributions of molecular gas masses are not significantly different with a median p-value = $0.05^{+0.21}_{-0.05}$.

\begin{figure}
    \centering
    \includegraphics[trim={ 0.2cm 0.2cm 0.25cm 0cm},clip, width=\linewidth]{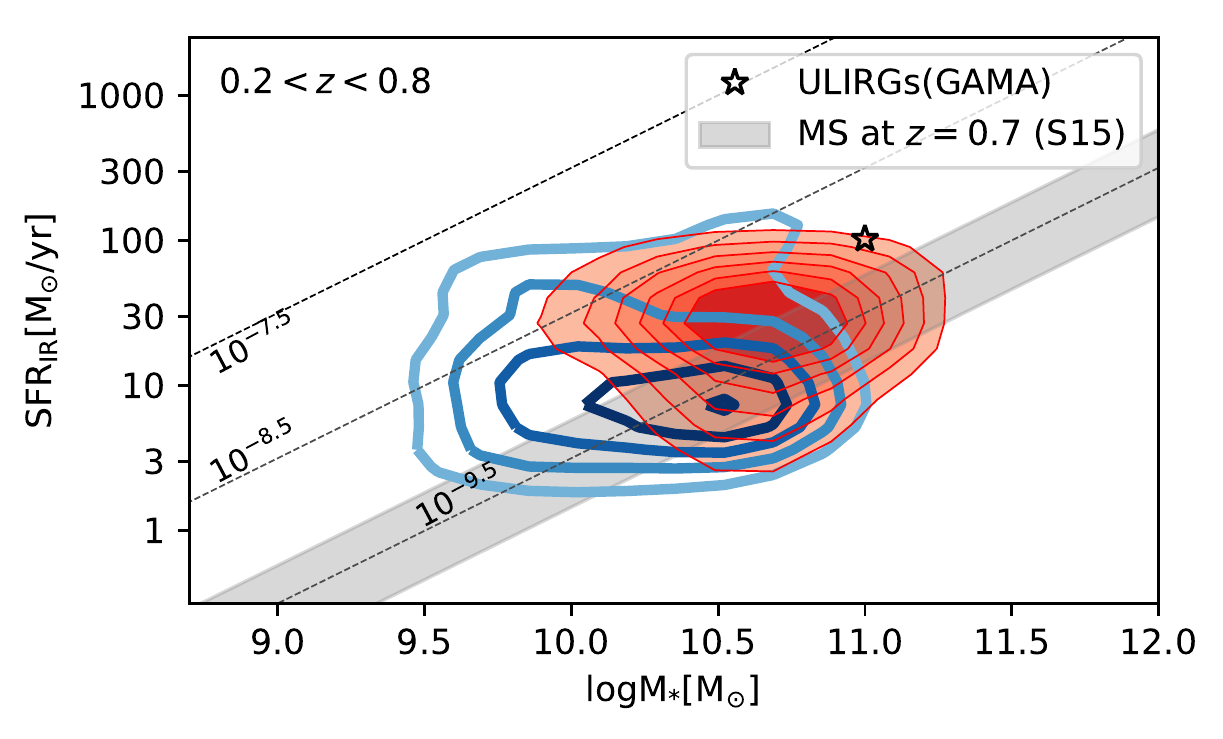}
    \caption{Composite two-dimensional posterior distributions of star-formation rates versus stellar mass for red QSOs (red shaded contour areas) and control QSOs (blue contour lines) for the low-$z$ sources. These are constructed by sampling 100 realization from the posterior of each source to properly account for uncertainties. Stellar masses and star-formation rates are estimated based on SED modelling. The distributions are consistent with the main sequence (grey shaded region), although as well with ULIRGS \citep[GAMA survey;][]{driver18} within the uncertainties. }
    \label{fig:MS}
\end{figure}

Despite this overall agreement of properties, we carry out a more focused analysis for the low-$z$ sample since we can assume the host galaxy properties are more reliably determined in these sources through the SED fitting. 
In Figure \ref{fig:MS} we plot the M$_{*}$-SFR relation for the red QSO sample in comparison to the luminosity-and-redshift-matched control sample for the redshift range $0.2<z<0.8$, where the overall SEDs had significant contribution from the host galaxy emission.
While both distributions are consistent with the main sequence and have similar sSFRs, the red QSOs are distributed in a more massive region ($\log \rm M_{*}(rQSO) = 10.49^{+0.43}_{-1.11}$ \Msun)\, than control QSOs ($\log \rm M_{*}(cQSO) = 10.14^{+0.60}_{-0.70}$ \Msun)\, although with considerable overlap.
The region occupied by red QSOs is consistent with that populated by ULIRGs at $z<1$ as shown by the median M$_{*}$-SFR relation of the GAMA Field ULIRGs \citep{driver18}.
However, this difference disappears if we use the M$_{\rm BH}$-based stellar mass estimates finding values of $\log  \rm M_{*}(rQSO) = 10.47^{+0.45}_{-0.59}$ \Msun\, and  $\log  \rm M_{*}(cQSO) = 10.44^{+0.54}_{-0.58}$ \Msun.
While this tentative finding is interesting, it is limited to this redshift bin (only 16 \% of the red QSO sample) and no conclusions can be drawn on the stellar masses of the general red QSO population based on this observation.

Overall, we find no strong differences in the host galaxy stellar populations and infrared properties between the red and control QSO populations.
In particular, the infrared luminosities from which star-formation rates were estimated and the molecular gas masses, as measured using a conversion from the 850 \micron\,luminosities, show equivalent distributions, suggesting no differences in the potential impact of AGN feedback on these properties in red or control QSOs.
These results suggest that there is no clear connection between the reddening properties of red QSOs and integrated properties at galaxy scales.
We discuss these points in more detail in Section \ref{sec:discussion}.

\section{Discussion}\label{sec:discussion}

We have studied the multiwavelength physical properties of red QSOs in comparison to a control QSO sample, carefully selected based on the observed $g^*-i^*$ colours from the SDSS survey photometry. 
Following the selection approach presented by \citet{klindt19}, we define our red and control QSO samples as the reddest 10 per cent, and the central 50 per cent of the colour distributions, respectively.
Motivated by fundamental differences found in the radio properties among these samples reported by \citet{klindt19, fawcett20, rosario20}, we undertook a comprehensive investigation of the FIR-to-UV emission to elucidate the nature of red QSOs. 
We utilised the Bayesian SED-fitting code \textsc{AGNfitter} to disentangle multiple physical components (i.e. the AGN accretion disk and dusty torus; the host-galaxy stellar and cold dust emission) responsible for the overall panchromatic emission over the FIR--UV waveband.
Overall, we found remarkable similarities in the SEDs between red and control QSOs which provide us with interesting insights on the nature of the reddening in red QSOs.
We discuss different implications of these results in the next sections.

Combining our SED estimates and the spectroscopic properties from SDSS QSO spectra, in Section \ref{subsec:intrinsicprops} we used the virial approximation to infer the distributions of black hole masses and Eddington ratios for our red and control QSO samples.
We found remarkably similar distributions and no significant differences in these properties.
Although this is in tension with a few results which relate red colours to different accretion properties \citep[e.g.][]{richards03, kim18} these literature results are based on small samples (e.g. few to handfuls of near-infrared-selected red QSO), and usually their comparison samples are subject to other selection strategies (e.g. optical selected) making direct comparisons challenging.
On the other hand, \gcr{the result that the reddening is linked to dust attenuation rather than intrinsic red QSO continua is in overall agreement with the majority of QSO investigations in the literature \citep[e.g.][]{richards03, glikman12, glikman13, kim18} and consistent with our detailed fitting to X-Shooter UV--near-IR spectroscopy for red and control QSOs selected in the same manner as this study (Fawcett et al. in prep.).}
However, although dust attenuation is the most straightforward explanation for the reddening in QSOs, the origin or distribution of the obscuring material is still unclear.
We investigate this question in the following sections.

\subsection{The origin of QSO reddening: No link with orientation and host galaxy ISM}

In this section we use our SED-fitting results to search for connections between the dust attenuation in red QSOs and the source of the attenuation; i.e. whether this occurs at nuclear or host-galaxy scales.
Firstly, we investigate whether the reddening in QSOs arise from specific configurations of the torus structure, which is in turn linked to viewing-angle and parametrised by a column density parameter in our dusty torus model.
We find the dusty `torus' components have almost identical shapes for both populations, with modelled column densities of $\log \rm N_{\rm H}\sim 21.5\, cm^{-2}$ in both cases (Figure \ref{fig:distributions_ebv_nh_r}) implying that no direct connection exists between the dust reddening in red QSOs and the emission from the dusty torus, responsible for the unification scheme.
On the contrary, in the context of unification, the tori modelled for the majority of red QSOs are consistent with a face-on unobscured view of prototypical Type 1 QSOs.
We believe systematics from our modelling technique are not affecting these results, given that this model has been shown to characterise inclination properties reliably, being able to robustly recover spectroscopic classifications of Type 1 and Type 2 AGN based solely on photometry \citep{CR16}.
This observation argues against the possible link between QSO reddening and inclination, and suggests the reddening must arise from other structures, potentially at different scales.

\gcr{Indeed, these results are supported by the observations presented by \citet{klindt19}, \citet{rosario20}, and \citet{fawcett20}, which find fundamental differences in the radio properties of red and control QSOs that are inconsistent with the scenario where the red QSOs are more inclined, on average, in comparison to the control QSOs. }
This finding is also in agreement with several recent studies of red QSOs that support an evolutionary scenario \citep{urrutia12, glikman12, banerji12, banerji17}, although in tension with the interpretation by \citet{wilkes02}, which used X-ray spectroscopy in red QSOs and interpret their values of $\log \rm N_{\rm H}(x-rays)\sim 21.5$ as a signature of intermediate inclination between Type 1 and Type 2.
We argue X-ray-based column densities are not necessarily linked to the torus structure, but might be connected to any gas structure on the line of sight, or different dust-to-gas ratios in the composition of the obscuring medium \citep{circosta19, lansbury20}. 

\begin{figure*}
    \centering
    \includegraphics[trim={ 0.2cm 0.2cm 0.3cm 0cm},clip, width=0.4\linewidth]{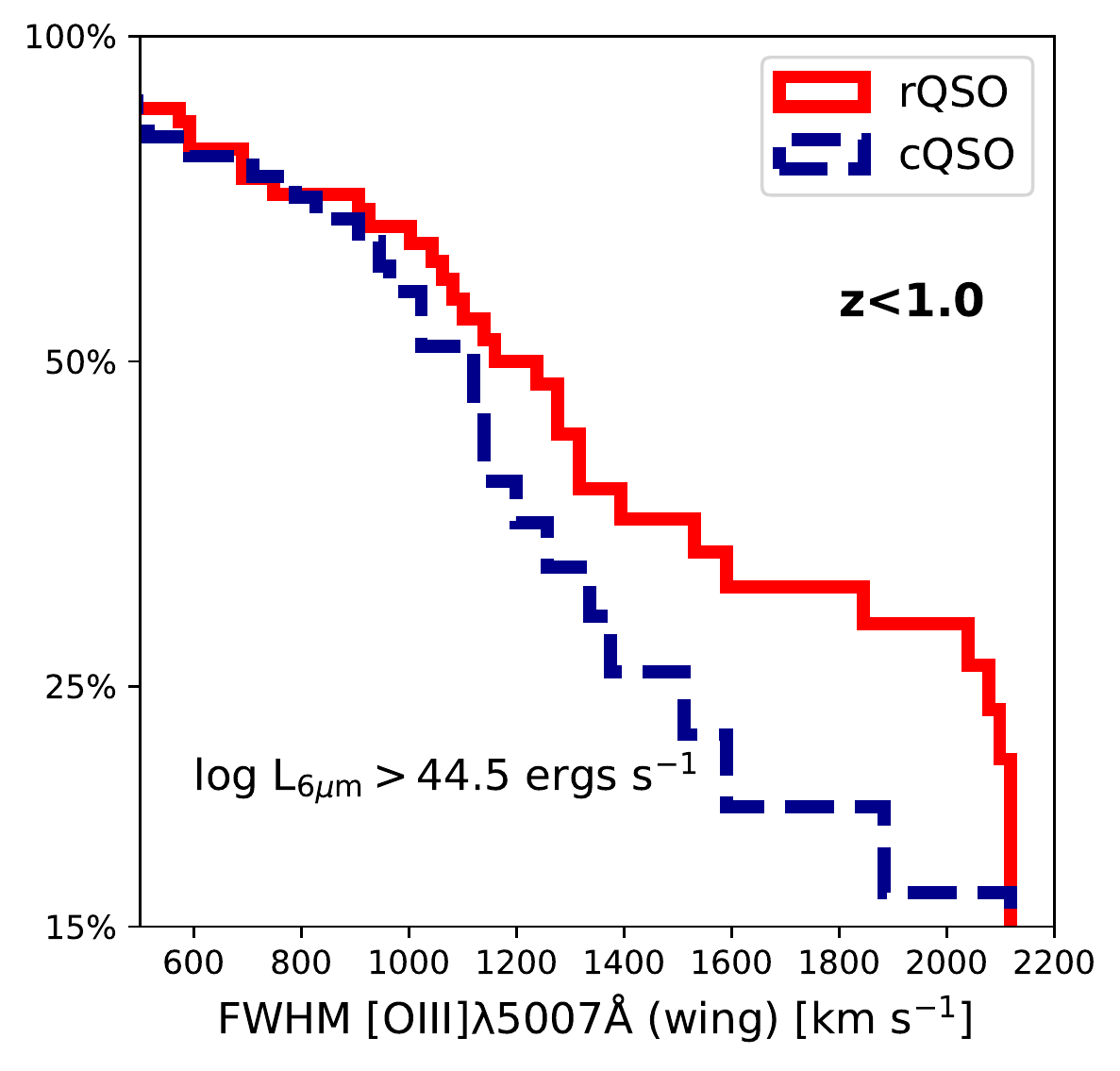}
    \includegraphics[trim={ 0.2cm 0.2cm 0.3cm 0cm},clip, width=0.4\linewidth]{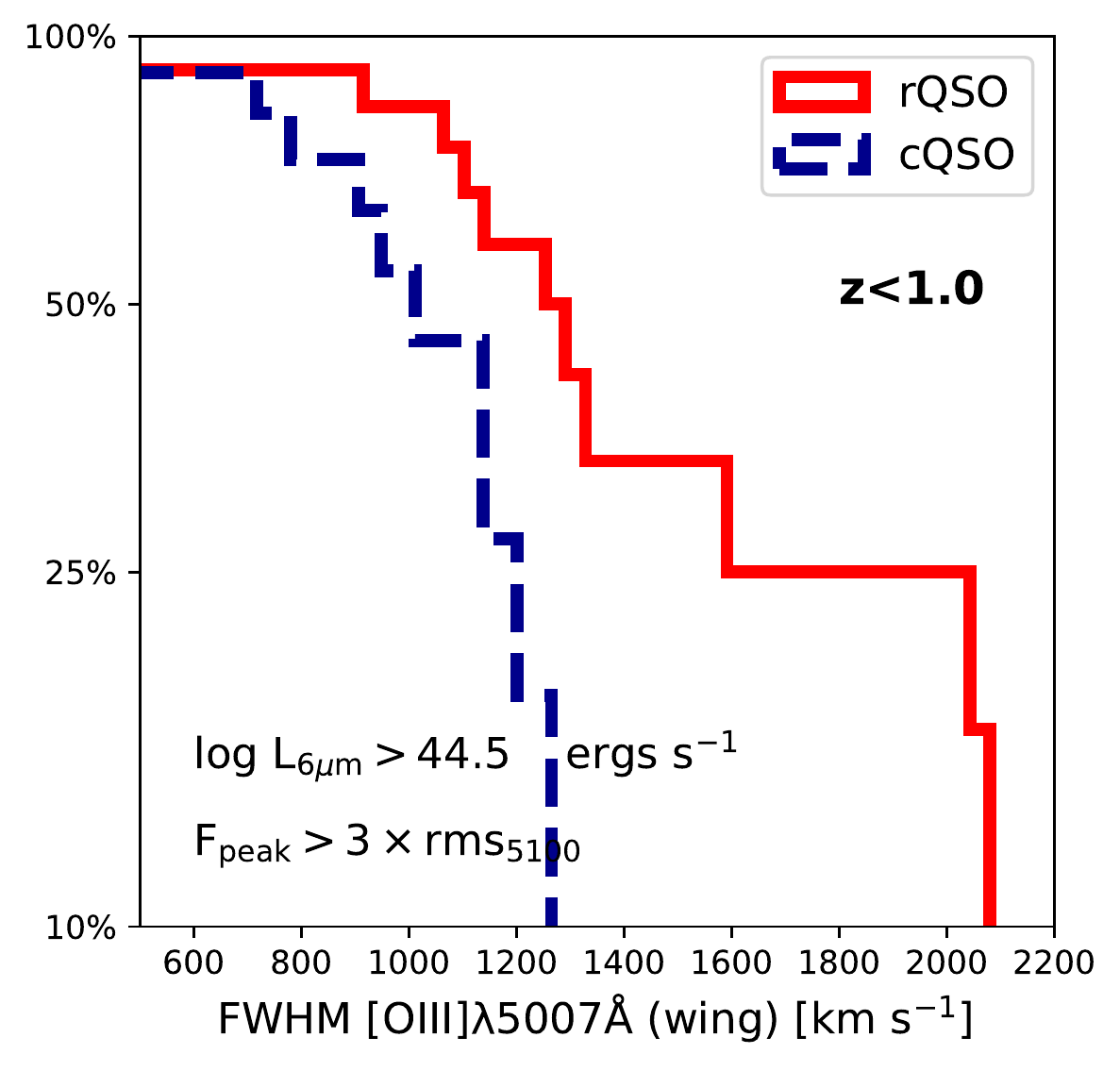}

    \caption{Cumulative distributions of the FWHMs (in the units of km s$^{-1}$) of the wing component of the [O~{\sc iii}]$\lambda 5007$ emission line for the red and control QSOs. Two cuts were required from our parent sample due to observability of the [O~{\sc iii}] line ($z<1$) and to avoid contamination from the host galaxy ($\log$ L$_{6 \mu \rm m} >44.5$). In the left panel the entire population is shown, without any further quality requirement. In the right panel, we restrict our study to high quality detections of the wing components, based on a SNR criterion defined in the plot, which is consistent with our visual inspection of the lines. The high-quality detection cut increases the evidence for a prevalence of high velocity winds in red QSOs as compared to control QSOs. } 
    \label{fig:OIII}
\end{figure*}

We further studied a possible link between the reddening and the torus reprocessing efficiency (Figure \ref{fig:RvsLbol}), and found the tori in red QSOs are at most equally efficient in reprocessing the intrinsic QSO emission, i.e. no additional dust-attenuation can occur in the torus structure.
This reinforces our first conclusion that no clear link is found between the nature of the tori in red QSOs and their colour.
In agreement with our results, \citet{kim18} present analysis of near-infrared spectra of 20 red QSOs  at $z<0.9$ finding no differences in their covering factors as compared to a larger samples of unobscured SDSS QSOs.
In conclusion, this set of evidence puts forward the scenario where no clear connection between reddening in QSOs and inclination exists, suggesting QSO reddening may be linked to a different source of obscuration, possibly related with evolution.

Our SED-fiting results have also allowed us to explore the connection between QSO reddening and the stellar masses, star-formation rates and interstellar medium of their host galaxies (Figure \ref{fig:distributions_hostgalaxy}), finding no significant differences in these properties between the red and control QSO populations.
In particular, we find that the molecular gas masses inferred from the rest-frame infrared 850 \micron \, emission, together with total infrared luminosities (used to estimate SFR) are equivalent for red and control QSOs.
We note that there are large uncertainties associated with the estimation of the rest-frame 850 \micron \, emission, since these are extrapolated from the inferred SEDs given our lack of photometric data coverage at those wavelengths (our reddest band is SPIRE 500 \micron).
However, these uncertainties are well constrained by our Bayesian inference approach.
We aim to investigate this point more in detail in a forthcoming publication based on ALMA 1\micron \, observations for a subset of red and control QSOs.
Based on our current results on the inferred stellar population and interstellar medium properties, we argue that no clear signature is found that the reddening might be linked to host galaxy scales.

\begin{figure}
    \centering
    \includegraphics[trim={ 0.6cm 0.2cm 0.35cm 0cm},clip, width=0.49\linewidth]{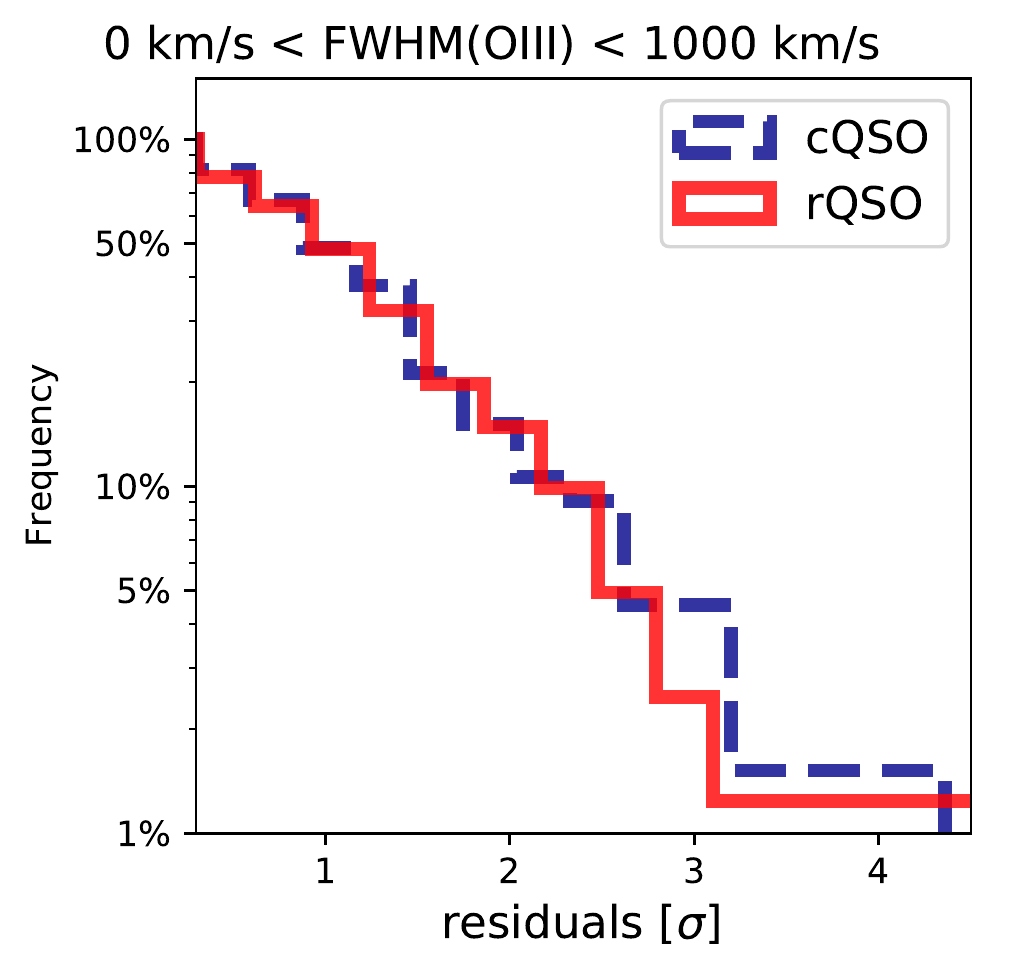}
    \includegraphics[trim={ 0.6cm 0.2cm 0.35cm 0cm},clip, width=0.49\linewidth]{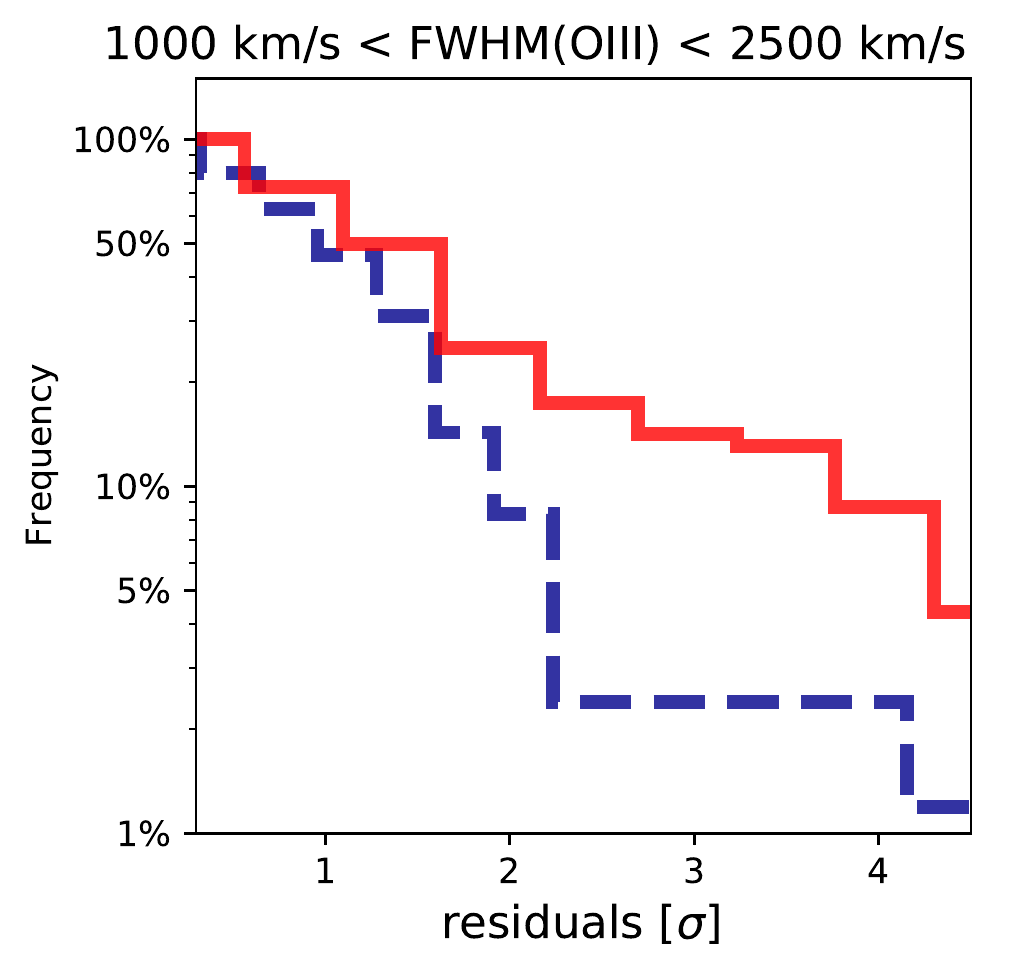}

    \caption{Cumulative distributions of the infrared residuals defined in Figure \ref{fig:residuals} for the red and control QSO samples, split into two bins of low-velocity (left panel) and high-velocity [O~{\sc iii}] wings (right panel). A clear predominance of residuals can be recognised for red QSOs as compared to control  QSOs in the right panel,  whereas  this difference disappears in  the  left  panel. This result supports the scenario where high-velocity winds, which are more prevalent in red QSO, are the main transport of obscuring medium in the QSO line of sight, producing the hot dust reprocessed emission and the colours of red QSOs.} 
    \label{fig:OIII-res}
\end{figure}
\subsection{The origin of QSO reddening: the dusty wind scenario}

To identify infrared features that are not fully recovered by the torus main structure, nor host galaxy emission, we searched for residuals from our SED-fitting of the red and control QSOs  (Figure \ref{fig:residuals}).
We find significant dust residuals evident as infrared bumps at $2-5$ \micron \, (consistent with hot dust at 500-1500 K), which are on average more present in red QSOs than in QSOs from the control sample. 
This trend is observed to be present across all redshift bins covered in our study, precluding any significant host galaxy contribution as the latter is negligible at higher redshifts and QSO luminosities.
Additionally, we find a general correlation between the prominence of these residuals and the QSOs reddening parameter E(B-V)$_{\rm BBB}$ (Figure \ref{fig:res_ebv}).
We interpret this observed infrared excess as an indicator of the nature of the obscuring medium, which causes the reddening in red QSOs.

Indeed, the temperatures indicated by these wavelengths can provide information on the scales over which the obscuring dust is distributed.
Following the prescription of \citet{barvainis87} \citep[similar to][]{kawaguchi10, stalevski16}, we assume a spatial distribution of dust as a flared disk, and we can estimate the radius where the dust needs to be located in order to produce this emission.
We estimate this following:
\begin{equation}
    \left(\dfrac{R}{\rm pc}\right) \simeq 1.3 \left(\dfrac{\rm L_{\rm bol}}{10^{46} \rm erg\, s^{-1}} \right) ^{0.5} \left(\dfrac{T_{\rm dust}}{1500\,\rm K}\right) ^{-2.8} \left(\dfrac{a}{0.05 \, \mu \rm m}\right)^{-0.5} 
\end{equation}
where we used the median bolometric luminosity values for the red QSO sample $\log L_{\rm bol}$= 46.5,  $T_{\rm dust}$ is the black body temperature corresponding the peak of the residual peak, which lies approximately at 1000 K, and $a$ is the dust grain size, for which we assume the Galactic ISM value of $\sim 0.05 \, \mu \rm m$.
We estimate that the dust producing the infrared excess emission is located at a radius of $R\sim 7$ pc from the central engine, which corresponds to a factor of 3-14 of the sublimation radius for average Galactic grains and large graphites, respectively.
Although we note these estimates are based on a very approximative approach, these values are consistent with the inner region of the torus.
Observationally, similar infrared excess or hot dust bumps around the same wavelengths have been previously found based on accurate infrared spectroscopy in nearby QSOs and Seyfert galaxies \citep{mor09, riffel09, burtscher15}, predominantly in unobscured, Type 1 AGN, local equivalents to those in our sample.
\gcr{In particular, \citet{mor09} analysed detailed infrared spectra of Palomar-Green (PG) QSOs finding that a hot dust black body component at temperatures 900-1800 K is required to fit the observed residuals. }
Despite not including optical data points, in contrast to our study, they are also able to reject any possible host galaxy contribution (see also \citealt{hernancaballero16}).
A similar infrared excess has been also reported in luminous Type 1 QSOs at intermediate redshifts ($z>1$), connecting it to emission arising from the sublimation radii of the torus \citep{temple20}.

While our infrared modelling describes only the torus emission, an additional emission component could be responsible for causing the reddening of the red QSOs.
One scenario which is being increasingly discussed in the observational and theoretical literature is the presence of dusty polar winds in AGN \citep{mor09, honig17, dipompeo18, baron19, stalevski19, venanzi20, harrison18, costa18}.
Indeed, the presence of winds at the scales estimated above, and the corresponding infrared emission at the wavelength where the infrared excess was found (2-5 $\mu$m) is predicted by theoretical work \citep[e.g.][]{hoenig19}, as a consequence of AGN and infrared radiation pressure leading to a puffed-up disk/wind-launching region \citep[e.g.][]{roth12}.
The scenario that the obscuring component is distributed in such a wind seems plausible according to recent work which suggests that dust `forms preferentially' in outflowing gas at these scales \citep{sarangi19}.
A widely-used tracer of winds and outflows in AGN is the [O~{\sc iii}]$\lambda$5007 emission line (from now on referred as [O~{\sc iii}]), which traces low-density warm ionised gas in the narrow-line region of AGN.
The presence of high velocities and strongly kinematically disturbed [O~{\sc iii}]-emitting gas is a sign-post for  outflows on scales of few pc to few kpc \citep[e.g.][]{baskin05, harrison14,  woo16, zakamska16}.
Since these scales are consistent with our estimation from the infrared residuals, we explore the potential presence of such winds and outflows in our data.

To examine the distribution of FWHMs of [O~{\sc iii}] wings, we use the spectral fitting output from the SDSS spectral catalogue by \cite{rakshit20} (partially through private communication).
The [O~{\sc iii}] line emission was covered by the SDSS spectral band only for sources at $z<1$.
A double Gaussian model was used to represent the [O~{\sc iii}] emission line; one for the core and another for the wing components.
During the fit, the FWHM of the core component was allowed to vary between 100 to 900 km s$^{-1}$ and the velocity offset was restricted to $\pm$ 1000 km s$^{-1}$ , while for the wing component, the FWHM was varied between 150 to 2100 km s$^{-1}$  and the velocity offset was restricted to $\pm$3000 km s$^{-1}$. 
To avoid possible contamination from host galaxy features, we make a further cut for QSO luminosities with $ \log$ L$_{6\mu \rm m}> 44.5$ \ergs.
We note that the luminosity cut and the redshift cut applied to cover the [O~{\sc iii}] emission reduce our original sample by 87 per cent (42 rQSOs and 31 cQSOs).
\gcr{In the left panel of Figure \ref{fig:OIII} we plot the distributions of [O~{\sc iii}] FWHM for the wing components of the red and control QSO samples for all sources with [O~{\sc iii}] measurements and luminosities of $\log \Lsixum> 44.5$ \ergs. }

We further test for reliability of the detection of the [O~{\sc iii}] wing components both through visual inspection and by comparing the significance of the emission line peak compared to the continuum noise at 5100$\, \AA$, defining the $\rm SNR_{OIII}=F^{[O~{\sc iii}]}_{\rm peak}/RMS_{5100}$ (following \citealt{rakshit18}).
Through visual inspection, we find this criterion is a good representation of the reliability of the wing.
After applying a reliability cut at $\rm SNR_{OIII}>3$ we remain with a sample of 23 sources (11 cQSOs, and 12 rQSOs).
The right panel of Figure \ref{fig:OIII} clearly shows the prevalence of the [O~{\sc iii}] wings in red QSOs as compared to the control QSOs.
In order to overcome the limitation of the small sample size and increase the statistical significance of our results, in Appendix \ref{appendix:OIII} we explore the [O~{\sc iii}] line emission properties for the red and control QSOs in the entire SDSS sky.
Using samples of $>1000$ QSO spectra we confirm the prevalence of higher [O~{\sc iii}] wing velocities in the overall red QSO populations as compared to the control sample with high significance (p-value$ =10^{-4}$).
The most striking difference lies in the fractions of red and control QSOs which fulfil the requirement of having [O~{\sc iii}] detections at all, or [O~{\sc iii}] measurements of a given $\rm SNR_{OIII}$.
Indeed, the comparison of the distribution of $\rm SNR_{OIII}$ for red and control QSO, matched in redshift and luminosity, is highly significant as well (p-value$ =10^{-27}$), as shown in Figure \ref{appendix:ALLSDSS}.

To make a more direct connection between the [O~{\sc iii}] high-velocity winds and the infrared residuals, in Figure \ref{fig:OIII-res} we show the cumulative distributions of the infrared residuals for all our QSO samples, divided into two bins of [O~{\sc iii}] line velocities in each panel.
In the high-velocity bin (right panel) a clear predominance of infrared residuals can be recognised for red QSOs compared to control QSOs, whereas in the low-velocity bin (left panel) the difference is negligible.
This result supports the scenario where high-velocity nuclear winds, which are more prevalent in red QSOs (Figure \ref{fig:OIII}), are the main transport of obscuring medium into the line of sight of the QSO, both producing the hot dust excess and the colours of red QSOs.

\begin{figure}
    \centering
    \includegraphics[trim={ 0.2cm 0.2cm 0.2cm 0cm},clip, width=0.95\linewidth]{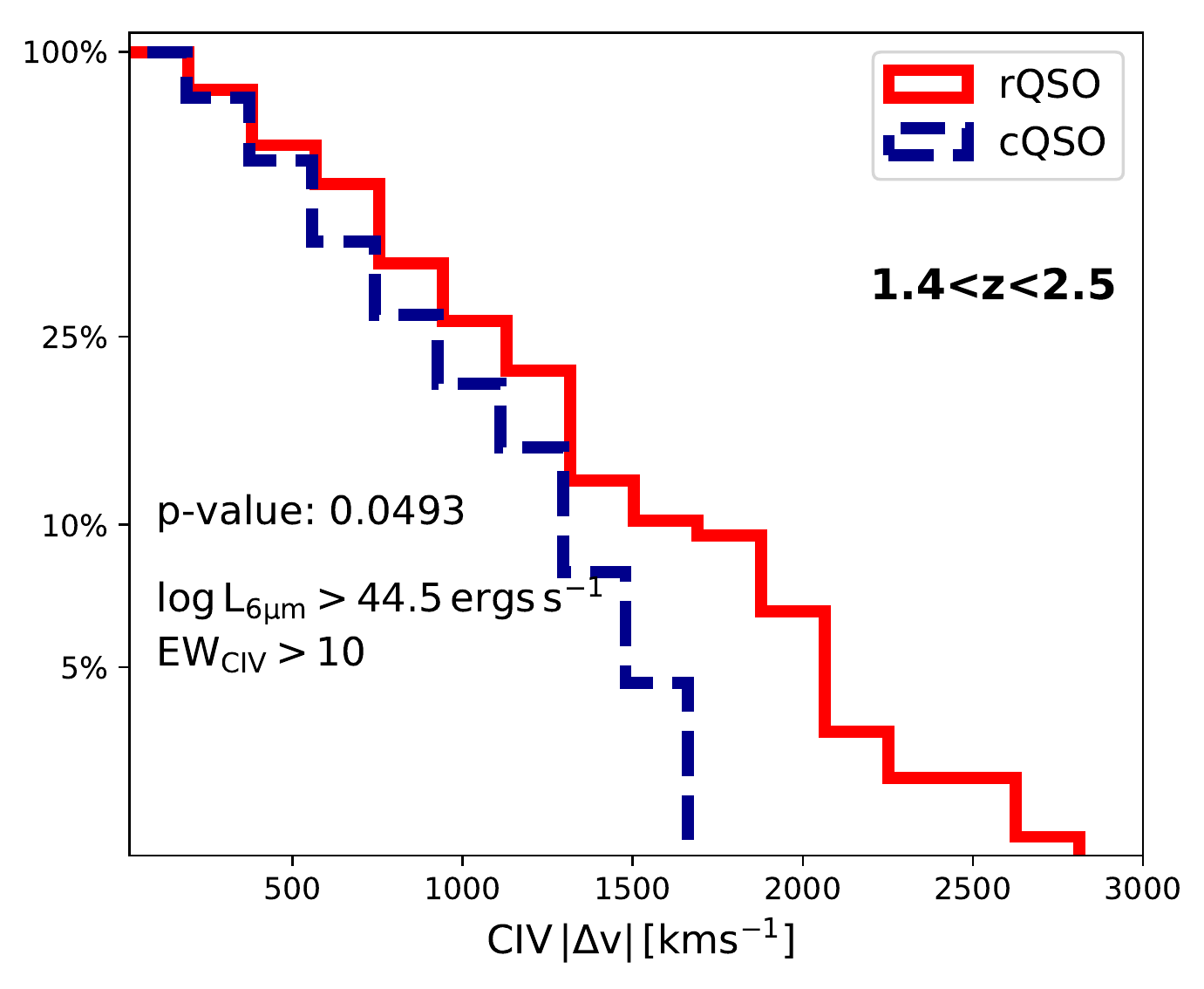}

    \caption{ \gcr{Cumulative distributions of the velocity shifts of the C {\sc iv} emission line (in units of km s$^{-1}$) for our samples of red and control QSOs, matched in redshift and luminosity. These samples cover the redshift range $1.4<z<2.5$, complementing the results presented using the [O {\sc iii}] lines at low redshift ($z<1$). A significant enhancement of C {\sc iv} velocity shifts is found for red QSOs as compared to control QSOs, which is consistent with our finding in Figure \ref{fig:OIII}. } }
    \label{fig:CIV}
\end{figure}
\begin{figure}
    \centering
    \includegraphics[trim={ 0.6cm 0.2cm 0.2cm 0cm},clip, width=0.48\linewidth]{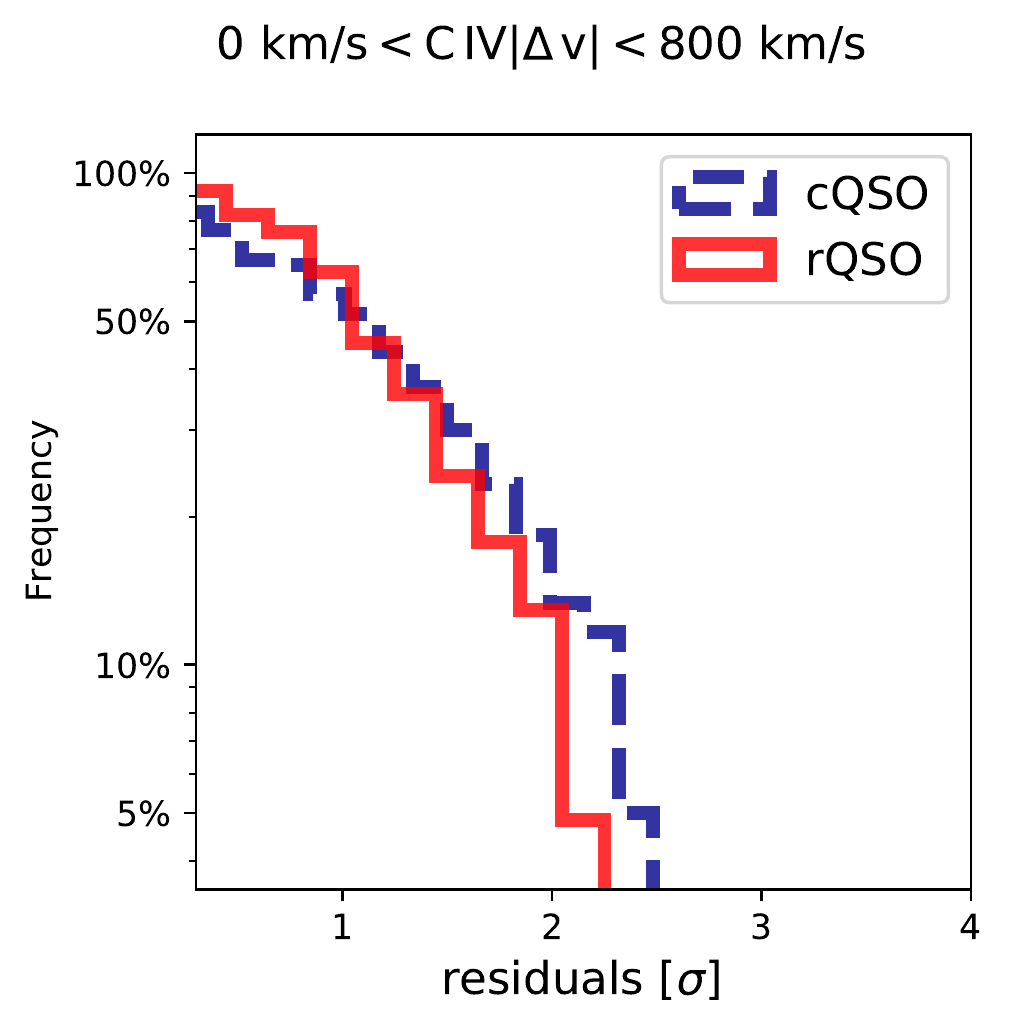}
    \includegraphics[trim={ 0.6cm 0.2cm 0.2cm 0cm},clip, width=0.48\linewidth]{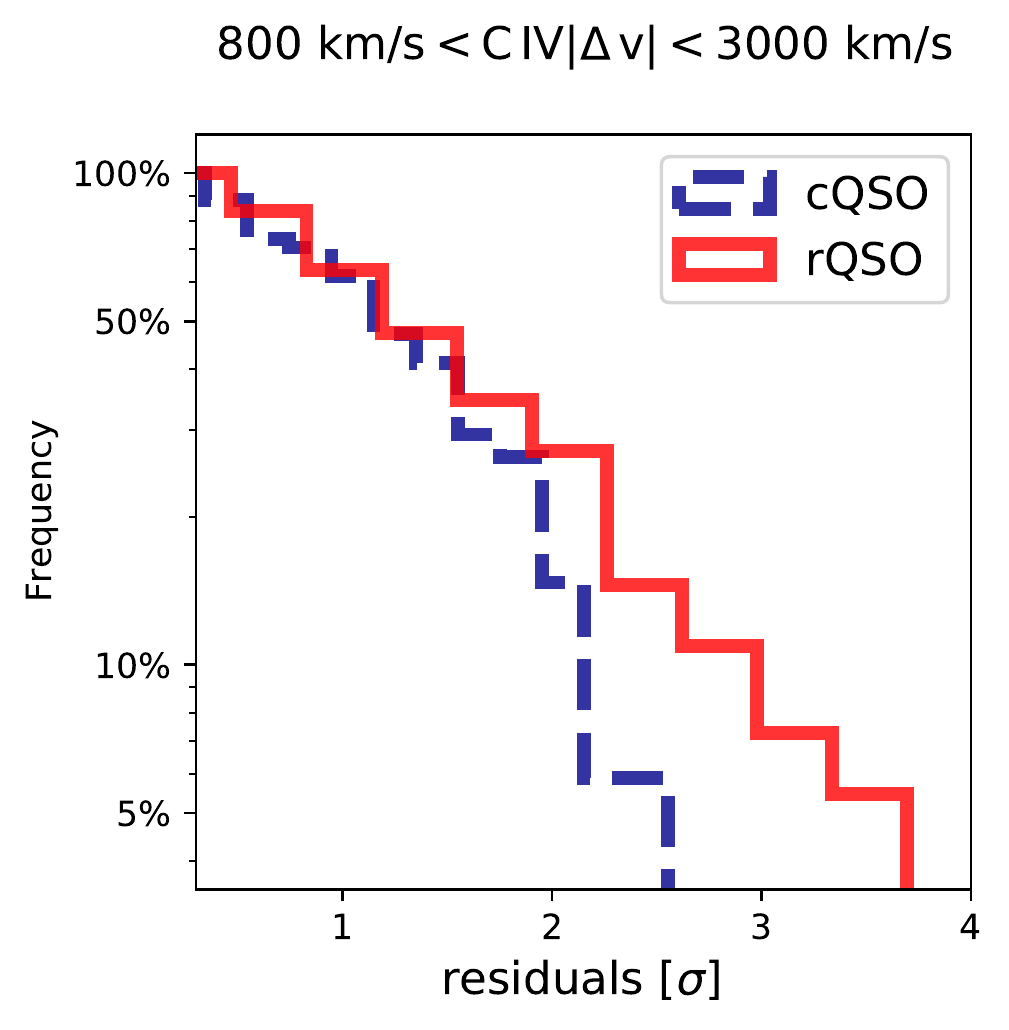}
    \caption{ \gcr{Cumulative distributions of the infrared residuals defined in Figure \ref{fig:residuals} for the red and control QSO samples, split into two bins of low (left panel) and high C~{\sc iv} velocity shifts (right panel). An enhancement of residuals can be seen for red QSOs with high C~{\sc iv} velocity shifts, a signature of potential winds or outflows.  This difference is not seen for control QSOs, which show overall a low incidence of residuals. These results are consistent with our [O~{\sc iii}] study at low redshifts (z<1), and support the proposed scenario of a connection between QSO reddening and the presence of dusty winds.} }
    \label{fig:CIV_residuals}
\end{figure}

\gcr{
Although these results provide significant evidence for the prevalence of high velocity winds in red QSOs, our conclusions are limited to low redshifts ($z<1$).
To rule out a redshift dependence of this result, we further investigate the presence of high-velocity winds in our red and control QSOs at $1.4<z<2.5$ by analysing the spectral properties of the C~\textsc{iv} emission line, reported by \citet{rakshit20}.
In particular, the C~{\sc iv} line profile is commonly used as a tracer of outflows that are driven off the accretion disk, as it often presents asymmetric shapes associated with high-velocity components. 
Such components are able to shift the observed velocities from the rest-frame velocity of the line by up to 5000 km\,s$^{-1}$ \citep[e.g.][]{richards11, coatman19, temple20}.  
We calculate the velocity shifts for our matched sample of red and control QSOs as $|\Delta v_{\text{C} \,\textsc{iv} }|= |c \times (\lambda_{\rm rest} - \lambda_{\rm peak})/ \lambda_{\rm rest}|$,  where $\lambda_{\rm rest}$ is the rest-frame wavelength  of the C \textsc{iv} doublet at $1549.48 \, \AA$, and $\lambda_{\rm peak}$ represent the peak wavelength measured and reported by \citet{rakshit20}. 
We remind the reader that the C~{\sc iv} emission line parameters in \citet{rakshit20} correspond to the entire C~{\sc iv} profile without subtracting a narrow component. This choice was made due to the ambiguity in the presence of narrow components in these lines \citep{shen11, shen20}.
A detailed description of the spectral fitting can be found in \citet{rakshit20}.
In addition to the redshift limits imposed to cover the C \textsc{iv} line in SDSS ($1.4<z<2.5$), we apply a cut in  luminosity ($\log \Lsixum > 44.5$ \ergs) for consistency with our [O \textsc{iii}] study and require a C \textsc{iv} line equivalent width with C \textsc{iv} EW$ >10\, \AA$ to ensure robust velocity shift estimates.}

\gcr{In Fig. \ref{fig:CIV}, we show the cumulative $|\Delta v|$ distributions for our luminosity-redshift matched sample.
We find that there is a significant difference in the distributions of |$\Delta$ v| of red and control QSOs, where red QSO show higher C \textsc{iv} velocity shifts.
In line with the study presented by \citet{coatman19}, where they find a correlation between the outflow signatures in the [O \textsc{iii}] and C \textsc{iv} emission lines, we conclude that this result is consistent with our findings for red and control QSOs at low $z$. 
This finding demonstrates that the enhanced velocities in the broad and narrow line region in red QSOs is a property which holds for both the low and high-$z$ population.
Finally, in Figure \ref{fig:CIV_residuals} we also investigate a direct relation between the high-velocity wind signature and the infrared residuals.
We find that red QSOs with higher C \textsc{iv} velocity shifts exhibit stronger hot dust excess, while no differences are found for the control QSOs irrespective of C \textsc{iv} spectral properties. 
These results are also consistent with what we observed using the [O~{\sc iii}] spectral properties, providing evidence of a connection between high-velocity dusty winds and QSO reddening across the entire redshift range probed ($0.2<z<1.$, and $1.4<z<2.5$). 
}

This result is the first study presenting a potential systematic connection between QSO optical reddening and the presence of dusty outflows.
However, it is in agreement with a large number of recent studies on luminous optical QSO samples \citep[e.g. ][]{mullaney13, bischetti17}, infrared \citep[e.g. ][]{dipompeo18}, as well as lower luminosity local AGN samples \citep[e.g. ][]{rojas20}.
Additionally, our result provides an interesting connection to extremely red QSOs (ERQs), where very high [O~{\sc iii}] velocities of $\rm FWHM >2000$ km s$^{-1}$ have been observed \citep{zakamska16, hamann17, mehdipour18, perrotta19, temple19, villar20}.
\gcr{In particular, \citet{zakamska16} studied 4 ERQs at $z=2.5$ and $\log \Lsixum>47$ \ergs\ finding luminous, broad (FWHM=2600-5000 km/s) and blue-shifted lines, estimating $\sim 3$ per cent of the bolometric luminosity of the QSOs was converted to outflow kinetic luminosity extended to a few kpcs.}
In line with our study, they found [O~{\sc iii}] kinematics positively correlated with infrared luminosities, although these four sources are extreme counterparts of our sample in terms of most properties presented here.  
A similar connection has been found by \citet{temple20}, reporting enhanced rest-frame 2 $\mu$m emission correlated with the blueshift of the [C IV]$\lambda 1550$ line for a sample selected in the optical and near-infrared bands.
In agreement with our findings, they find that objects with stronger signatures of nuclear outflows tend to have a larger covering fraction of dust around the sublimation-temperature, although they find the sources with higher excess have bluer continua and argue that no connection exists to QSO reddening.
While several differences exist between our sample selection as they do not focus on a colour selection, this result is intriguing and underlines the need for more detailed spectroscopic investigations both in the optical and mid-infrared regimes to confirm our interpretation.

\subsection{On the role of red QSOs in galaxy evolution} \label{subsec:conclusion_galaxy}

The host galaxies of red QSOs have been explored for just a few objects, given the difficulty in disentangling the stellar emission from the dominant accretion disk component and the lack of homogeneous deep Herschel or ALMA imaging.
Indeed, due to these challenges existing studies find diverging evidence regarding several questions around the host galaxies, such as whether the host galaxies of red QSOs are undergoing mergers in a similar manner \citep{zakamska19} or more often than blue QSOs \citep{urrutia08, urrutia12}, and whether red QSOs host galaxies are more strongly star-forming than typical QSOs \citep{georgakakis09} or whether there are no significant differences \citep{urrutia12, wethers20}. 
Two main challenges in the task of comparing these different studies is that they often involve low statistics and inhomogeneous selection effects.

In our work we exploit the availability of rich multiwavelength UV-to-FIR ancillary data and the capability of \textsc{AGNfitter} to robustly infer underlying uncertainties, in order to model the host galaxy properties of our controlled samples of red and control QSOs.
Applying our careful comparative exercise on a statistical sample, we find no significant differences in the stellar mass, star-formation rates, and molecular gas estimates of red and control QSOs (Figure \ref{fig:distributions_hostgalaxy}). 
Taking into account the uncertainties associated with our measurements, we find both populations are overall consistent with the massive end of the main sequence of star forming galaxies, in line with previous studies of the star formation in general samples of optical QSOs \citep{rosario13, stanley17, wethers18}.
Interestingly, we infer no significant differences in the molecular gas masses of the red and control QSOs.

Since our SED-based results argue against a connection between the reddening in QSOs and inclination of the dusty torus, we  further evaluate our observations in the context of the galaxy evolution models \citep{sander88, hopkins08, narayanan10} that propose red QSOs as intermediate stages between dusty star forming galaxies, and UV-bright unobscured QSOs.
Based on the similarities found in the host-galaxy properties of red and control QSOs, we conclude that while an evolutionary link might exist between red and control QSOs, the impact it has on the host galaxies is not large enough for it to be apparent in our statistical study.
This could occur if the timescales separating these phases are not long enough to produce such an effect.
To put this into context, recent simulations by \citet{costa20} indicate small-scale AGN winds would require timescales of $\sim 100\rm\,Myr-1\,Gyr$ to significantly reduce star formation in their host galaxies, through the removal of high-density gas or suppression of halo gas accretion when coupled to large-scale outflows.
By comparison, the typical estimated lifetime for a QSO is $\sim 10$ Myrs \citep[e.g.][]{martini01}. 
Assuming all QSOs undergo a red QSO phase then the lifetime for this phase on the basis of our sample would be ~1 Myr (i.e. red QSOs comprise 10\% of our overall QSO sample). 
We therefore conclude that the QSO lifetime is too brief to be able to expect significant differences in the host galaxies between red and control QSOs.
We note though that given our uncertainties, subtle changes in these properties might be still present.

In the context of the proposed evolutionary sequence between red and control QSO, we speculate that the presence of these winds could be a key ingredient for the evolution. 
In such a scenario, these small-scale dusty winds would redden the QSO continuum, while expelling a fraction of the gas and dust content from the central regions, becoming progressively less energetic with less available material.
This would be in line with models \citep[e.g. ][]{roth12} which suggest that powerful small-scale winds would accelerate through radiation pressure, which within a given timescale, would be efficient in unobscuring the AGN, with limited impact on the host galaxy \citep[e.g. ][]{costa18, costa18a}. 
The evidence that the infrared residuals correlate on average with the E(B-V) parameter (estimated independently from the optical SED), and the fact that a fraction of control QSOs also present high-velocity winds, is consistent with the scenario of a potential transition between these populations.

\section{Conclusions and summary}

Motivated by recent evidence of fundamental differences in the radio properties of red and control QSO, we presented a multiwavelength comparative analysis of $\sim$1800 red and control SDSS QSOs, where red were defined as the 10 per cent reddest, and control as the central 50 per cent of the population in the ($g^*-i^*$) colour distribution.
Using the Bayesian MCMC-based \textsc{AGNfitter} code, we performed a detailed AGN-tailored SED-fitting analysis on a rich compilation of multiwalength photo\-metry, covering up to 29 photometric bands from the UV to the far-infrared.
The SED fitting results together with spectroscopic information extracted from the QSO spectra led us to the following conclusions:

\begin{enumerate}[label=(\roman*)]

    \item  \textsc{AGNfitter} models the dust attenuation of the accretion disk emission using an SMC reddening law, finding distributions of reddening of E(B-V)$_{\rm BBB} = 0.02^{+0.04}_{-0.01}$ mag (A$_{\rm V} = 0.05^{+0.11}_{-0.03}$ mag) for our control sample of `average' optically-selected QSOs. 
    Our red QSO selection has median reddening values of E(B-V)$_{\rm BBB} = 0.12^{+0.21}_{-0.08}$ (A$_{\rm V} = 0.32^{+0.57}_{-0.21}$ mag), demonstrating that we are probing \textit{mild attenuation in the most reddened optically-selected QSO}. 
    Already with this mild attenuation, fundamental differences have been previously found in the radio properties of these sources \citep{klindt19,rosario20,fawcett20}.
    \item Apart from the diversity in attenuation of the accretion disk emission, we find \textit{no strong variations in the multiwavelength SEDs} of red and control QSOs across all optical-to-FIR wavelengths. \gcr{Based on spectral properties and reddening-corrected continuum measurements, we evaluate intrinsic properties of our QSO samples, such as black hole masses and Eddington ratios, finding \textit{no significant differences in the accretion properties} of the two populations.}
    \item We find the bulk of the infrared emission of both red and control QSOs are well described with the same torus emission structure, associated with low column densities of $\log N_H<22$ and face-on viewing angle. Additionally, we compare the torus reprocessing efficiency ($\rm L_{\rm tor}/L_{\rm bol}$) and covering factors, finding that the red QSO population has covering factors equivalent to those of control QSOs. The similarity of the infrared emission of the torus component suggests that there is \textit{no connection between the QSO reddening and the emission expected from the AGN torus}, disproving unification models as the origin of the red QSO nature.
    \item We investigate the presence of residual emission from infrared components other than the torus. Interestingly, we find a \textit{higher incidence of residual flux in the rest-frame $2-5$ \micron \, SEDs of red QSO}, associated with temperatures of $500-1700$ K and locations of $\sim 5-10$ pc from the nucleus. We suggest the origin of this emission lies in nuclear high-velocity winds, for which we find tentative evidence through high-velocity ionised winds \gcr{(traced by broad [O~{\sc iii}] line wings of $\rm FWHM > 1000$ $\rm km \, s^{-1}$ at low-$z$, and C~{\sc iv} velocity shifts at high-$z$)}.  In particular, we find a \textit{higher incidence of high velocity winds for red QSOs} as compared to control QSOs, and more prominently when residual fluxes are present. 
    \item We measure the host galaxy properties of red and control QSOs. 
    We find \textit{no significant differences in the stellar masses or star-formation rates} of red and control QSOs. Interestingly, we find no significant differences in the masses of the molecular gas reservoirs of the two populations, as inferred from extrapolations of the cold dust SEDs. We conclude no clear link can be recognised between the host galaxy properties and the reddening of QSOs, suggesting that \textit{the driver of the differences between red and control QSOs must be on nuclear scales}.

\end{enumerate}

Based on our statistical comparative study, we find tentative evidence for a connection between the reddening in red QSOs and the presence of dusty winds in the QSO nuclear region. Within the QSO evolutionary framework, high-velocity winds can indeed play an important role in ejecting the material away from the nuclei, consistent with observations that red QSOs are usually more abundant at higher bolometric luminosities.
In a forthcoming paper we connect our UV-to-FIR findings with deep low-frequency radio observations to expand our understanding of the nature of red QSOs and the origin of their enhanced radio detections.
\gcr{More detailed observations of the infrared spectra of a controlled QSO sample are required in order to confirm the prevalence of the hot dust emission component around $\sim 2-5\,\mu$m in red QSOs and its link to dusty winds, as the origin of QSO reddening.}

\section*{ACKNOWLEDGEMENTS}
\gcr{We would like to thank the anonymous referee for their insightful comments and suggestions which have considerably improved the manuscript.}
We would like to thank Dominika Wylezalek and George Lansbury for discussions that contributed to the analysis and interpretation of our results.
G.C.R. acknowledges the ESO Fellowship Program and a Gruber Foundation Fellowship grant sponsored by the Gruber Foundation and the International Astronomical Union.
D.M.A and D.J.R. gratefully acknowledge support from the Science Technology Facilities Council (STFC) through grant ST/000244/1.
M.S. acknowledges support by the Ministry of Education, Science and Technological Development of the Republic of Serbia through the contract no. 451-03-68/2020-14/200002 and the Science Fund of the Republic of Serbia, PROMIS 6060916, BOWIE. 
V.A.F. acknowledges a quota studentship funded by the STFC.
P.N.B. is grateful for support from the UK STFC via grant ST/R000972/1.
M.B. acknowledges support from INAF under PRIN SKA/CTA FORECaST and from the Ministero degli Affari Esteri della Cooperazione Internazionale - Direzione Generale per la Promozione del Sistema Paese Progetto di Grande Rilevanza ZA18GR02.
R.K. acknowledges support from the STFC through an STFC studentship via grant ST/R504737/1.
The analysis in this work made use of AstroPy \citep{astropy}, Matplotlib \citep{matplotlib}, SciPy \citep{scipy} and Pandas \citep{pandas}.

Funding for the SDSS and SDSS-II has been provided
by the Alfred P. Sloan Foundation, the Participating Institutions, the National Science Foundation, the U.S. Department of Energy, the National Aeronautics and Space Administration, the Japanese Monbukagakusho, the Max Planck Society, and the Higher Education Funding Council for England. The SDSS Web Site is http://www.sdss.org/.
The SDSS is managed by the Astrophysical Research Consortium for the Participating Institutions. The Participating Institutions are the American Museum of Natural History, Astrophysical Institute Potsdam, University of Basel, University of Cambridge, Case Western Reserve University, University of Chicago, Drexel University, Fermilab, the Institute for Advanced Study, the Japan Participation Group, Johns Hopkins University, the Joint Institute for Nuclear Astrophysics, the Kavli Institute for Particle Astrophysics and Cosmology, the Korean Scientist Group, the Chinese Academy of Sciences (LAMOST), Los Alamos National Laboratory, the Max-Planck-Institute for Astronomy (MPIA), the Max-Planck-Institute for Astrophysics (MPA), New Mexico State University, Ohio State University, University of Pittsburgh, University of Portsmouth, Princeton University, the United States Naval Observatory, and the University of Washington. This publication makes use of data products from the Wide-field Infrared Survey Explorer, which is a joint project of the University of California, Los Angeles, and the Jet Propulsion Laboratory/California Institute of Technology, funded by the National Aeronautics and Space Administration.
The Herschel Extragalactic Legacy Project, (HELP), is a European Commission Research Executive Agency funded project under the SP1-Cooperation, Collaborative project, Small or medium-scale focused research project, FP7-SPACE-2013-1 scheme, Grant Agreement Number 607254.
\bibliographystyle{aa}
\bibliography{mwQSO}
\begin{appendix} 

\section{[O~{\sc iii}] wing properties of red QSOs and control QSOs in the SDSS survey}\label{appendix:OIII}

\begin{figure}
    \centering
    \includegraphics[trim={ 0.2cm 0.2cm 0.2cm 0.2cm},clip, width=0.62\linewidth]{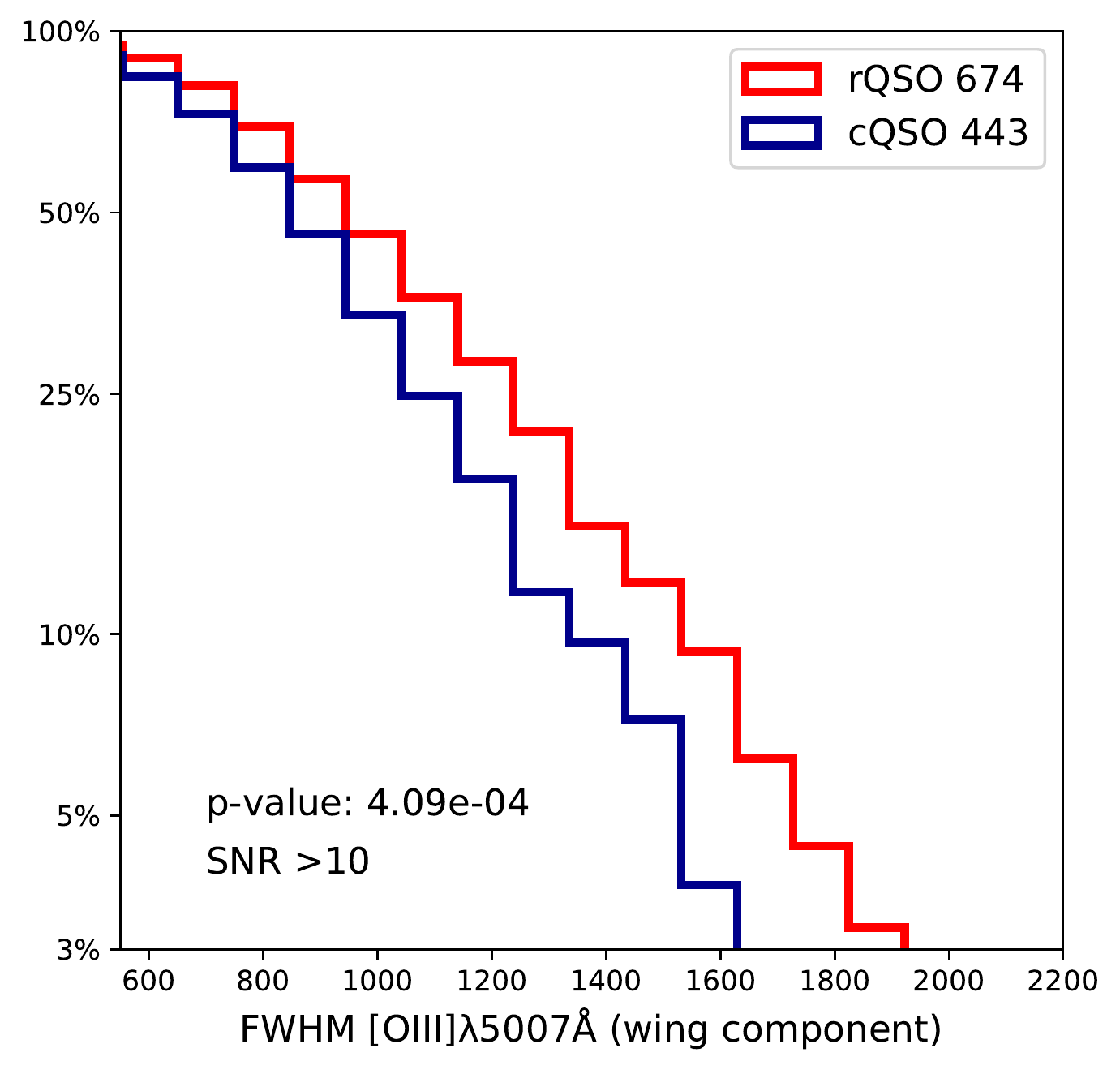}
    \includegraphics[trim={ 0.2cm 0.2cm 0.3cm 0.2cm},clip, width=0.62
\linewidth]{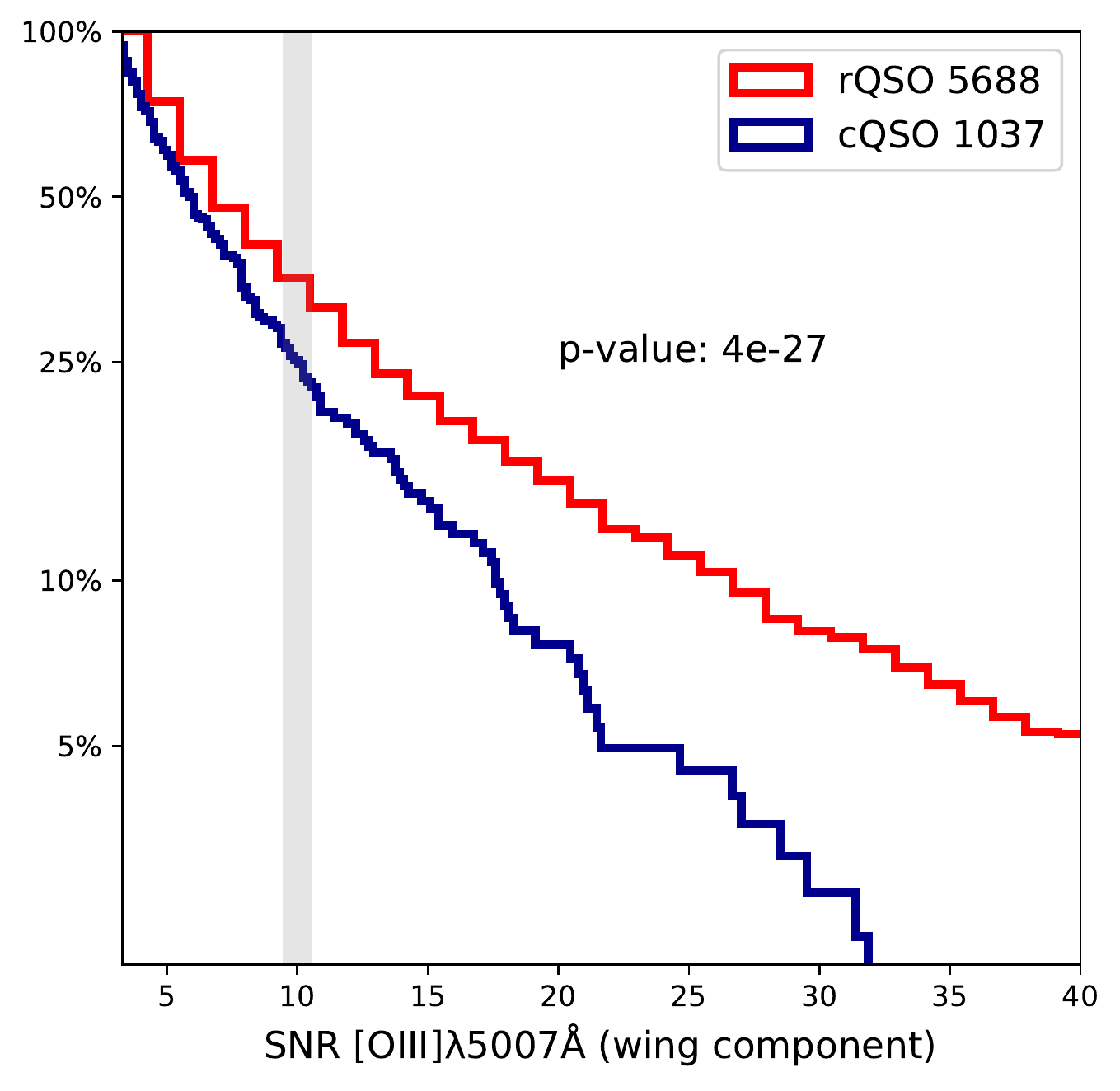}

    \caption{[O~{\sc iii}] emission line properties for the luminosity-and-redshift-matched red and control QSO samples from the entire SDSS survey. The upper panel shows the comparison of all red and control QSOs with [O~{\sc iii}] line measurements with F$_{\rm peak}> 10 \times \rm rms_{5100\AA}$. It is important to note the number of red QSOs satisfying the condition is $\sim 50$ per cent larger than the control QSOs. \gcr{This is shown more clearly in the lower panel, where the distribution of signal-to-noise ratio of the wing emission-line component, defined as = F$_{\rm peak}\rm (wing)/ rms_{5100\AA}$, is shown for red and control QSOs.} While here the only condition for a source to be included  was to have [O~{\sc iii}] detections, the number of red QSOs satisfying this condition is 5 times larger than that of the control sample. } 
    \label{appendix:ALLSDSS}
\end{figure}

To complement the analysis of the [O~{\sc iii}] wing properties of our red and control QSO sample we reapply our sample selection to the entire sky coverage of the SDSS survey.
In this way we also overcome the limitation of the small sample statistics discussed in Section \ref{sec:discussion} after the consideration of signal-to-noise ratio cuts for the [O~{\sc iii}] measurements.
\gcr{After applying our selection (WISE detected sources, classified based on their $g^{*}$-$i^{*}$ colours) and a redshift cut to $z<1$ to ensure the [O~{\sc iii}] is covered by the SDSS spectra, overall we find a total of 6753 red QSOs and 6753 control QSOs, matched in redshift and luminosity.}
We then compare the signal-to-noise ratio of the [O~{\sc iii}] measurements of this sample as shown in the lower panel of Figure \ref{appendix:ALLSDSS}, finding a significant difference in the signal to noise ratio of the [O~{\sc iii}] emission in red QSOs.
Interestingly, a larger fraction of control QSOs was discarded since these did not fulfil the requirement of having [O~{\sc iii}] detections. 
We also investigate the difference in the FWHM of the [O~{\sc iii}] wings, finding a significant enhancement of high-velocity ionised outflows in red QSOs, confirming the measurements discussed in Section \ref{sec:discussion} for a subset of these.

\end{appendix}
\end{document}